\documentclass[fleqn,usenatbib]{mnras}

\usepackage{newtxtext,newtxmath}

\usepackage[T1]{fontenc}

\DeclareRobustCommand{\VAN}[3]{#2}
\let\VANthebibliography\thebibliography
\def\thebibliography{\DeclareRobustCommand{\VAN}[3]{##3}\VANthebibliography}

\usepackage[tight,footnotesize]{subfigure}
\usepackage{color}
\usepackage{graphicx}	
\usepackage{enumitem}
\usepackage[dvipsnames]{xcolor}
\usepackage{colortbl}

\usepackage{dcolumn}
\usepackage{bbm}
\usepackage{latexsym}
\usepackage{gensymb} 

\usepackage{xfrac}
\usepackage{calc}
\usepackage{longtable}
\usepackage{dashrule}
\usepackage{multirow}%
\usepackage{rotating}
\usepackage{adjustbox}
\usepackage[graphicx]{realboxes}
\usepackage{pdflscape}
\usepackage{ragged2e}

\usepackage{subcaption}

\usepackage{bm}

\usepackage{hyperref}
\hypersetup{colorlinks=true,citecolor=blue,linkcolor=blue}
\usepackage[normalem]{ulem}

\usepackage{amsmath}
\usepackage{natbib}

\usepackage{float}
\usepackage{perpage}
\usepackage{morefloats}

\usepackage{comment}

\def\aap{A\&A}

\def\apj{ApJ}
\def\apjl{ApJL}
\def\apjs{ApJS}
\def\apss{Astrophysics and Space Science}
\def\jcap{JCAP}
\def\mnras{MNRAS}
\def\na{New Astronomy}

\def\physrep{Physics Reports}

\definecolor{Gray}{gray}{0.85}
\definecolor{PasteYellow}{rgb}{0.99,0.99,0.59}
\definecolor{LightRed}{rgb}{1,0.80,0.79}
\definecolor{LightCyan}{rgb}{0.88,1,1}
\definecolor{LightBlue}{rgb}{0.667, 0.847, 1}
\definecolor{LightGreen}{rgb}{0.564705882, 0.933333333, 0.564705882}

\definecolor{ElectricLime}{rgb}{0.658823529, 1.0, 0.0156862745}
\definecolor{FreshGreen}{rgb}{0.411764706, 0.847058824, 0.309803922}
\definecolor{LightEggplant}{rgb}{0.537254902, 0.270588235, 0.521568627}
\definecolor{NastyGreen}{rgb}{0.439215686, 0.698039216, 0.247058824}

\makeatletter
\DeclareRobustCommand{\textsupsub}[2]{{%
		\m@th\ensuremath{%
			^{\mbox{\fontsize\sf@size\z@{##1}}}%
			_{\mbox{\fontsize\sf@size\z@{##2}}}%
		}%
}}
\makeatother

\title[Hubble tension in an anisotropic Universe]{Hubble tension in an anisotropic Universe}

\author[M. Deliyergiyev et al.]{
Maksym Deliyergiyev,$^{1}$\thanks{E-mail: maksym.deliyergiyev@hlrs.de  (MD)}
Morgan Le Delliou,$^{2,3,4,5,6}$\thanks{Corresponding Author: delliou@lzu.edu.cn,Morgan.LeDelliou.IFT@gmail.com  (MLeD)}
and Antonino Del Popolo$^{7}$\thanks{E-mail: adelpopolo@oact.inaf.it  (ADP)}
\\
$^{1}$High Performance Computing Center Stuttgart (HLRS), Universität Stuttgart, 70550 Stuttgart, Germany\\
$^{2}$Institute of Theoretical Physics \& Research Center of Gravitation, Lanzhou University, Lanzhou 730000, China\\
$^{3}$Key Laboratory of Quantum Theory and Applications of MoE, Lanzhou University, Lanzhou 730000, China\\
$^{4}$Lanzhou Center for Theoretical Physics \& Key Laboratory of Theoretical Physics of Gansu Province, Lanzhou University, Lanzhou 730000, China\\
$^{5}$Instituto de Astrofísica e Ciências do Espaço, Universidade de Lisboa, Campo Grande, 1769-016 Lisboa, Portugal\\
$^{6}$Université de Paris-Cité, APC-Astroparticule et Cosmologie (UMR-CNRS 7164), F-75006 Paris, France\\
$^{7}$Dipartimento di Fisica e Astronomia, University of Catania, Viale Andrea Doria 6, 95125 Catania, Italy
}

\date{Accepted XXX. Received YYY; in original form ZZZ}

\pubyear{\the\year{}}

\begin{document}
\label{firstpage}
\pagerange{\pageref{firstpage}--\pageref{lastpage}}
\maketitle

\begin{abstract}
We explore the Hubble tension within an anisotropic cosmological framework by revisiting the Bianchi type-I model introduced in Le Delliou {\it et al.} 2020. Motivated by ongoing debates surrounding back-reaction effects and observed anomalies in the cosmic microwave background (CMB), we investigate whether a departure from isotropy in the late Universe could reconcile the observed discrepancies in Hubble constant measurements.
Using a Bayesian inference framework, we constrain the model parameters employing multiple nested sampling algorithms: \texttt{bilby}, \texttt{PyMultiNest}, and \texttt{nessai}. We perform the analysis under both uniform and Gaussian priors, allowing us to systematically assess the sensitivity of the inferred cosmological parameters to different prior assumptions. This dual-prior strategy balances agnostic parameter exploration with constraints informed by theory and observation.
Our findings demonstrate the reliability of our inference pipeline across different samplers and emphasize the crucial role of prior selection in non-standard cosmological model testing. The results suggest that anisotropic models remain viable contenders in addressing current cosmological tensions: even though the present model does not show alleviation of the Hubble tension, the data points towards anisotropies.
Future work may extend this methodology to more complex anisotropic scenarios and incorporate additional cosmological probes such as CMB polarization and gravitational wave standard sirens.
\end{abstract}



\section{Introduction}
\label{sec:Intro}

The $\Lambda$CDM model, in the past decades, has successfully passed many tests \citep{Komatsu2011,Planck2014_XVI,
DelPopolo2013}. This model is based on several assumptions. One of the assumption is that General Relativity (GR) is the correct theory of gravitation. Unfortunately, if the gravity is correctly described by GR, observations indicate the existence of a larger content of mass-energy than predicted \citep{DelPopolo2007,DelPopolo2014a,Bull2016}.
Non-baryonic and non-relativistic particles dominate the mass-energy of the universe, indicated as "cold dark matter" \citep{DelPopolo2014}, and a second component, dubbed "dark energy" (DE), a fluid with exotic properties such as negative pressure giving rise to the accelerated expansion of the universe, is needed to correctly explain the observations. In its simplest form, the $\Lambda$CDM model, DE is represented by the cosmological constant 
$\Lambda$. As shown in several papers, and using precision data \citep{Spergel2003,Komatsu2011,DelPopolo2007}, the $\Lambda$CDM model reveals some drawbacks and tensions both at large scales \citep{Eriksen2004,Schwarz2004,Cruz2005,Copi2006,Macaulay2013,Planck2014_XVI,Raveri2016}, and at small scales 
\citep{Moore1999,deBlok2010,Ostriker2003,BoylanKolchin2011,DelPopolo2014a,DelPopolo2014d,DelPopolo2017a}. 
Despite a large campaign of direct and indirect searches \citep{Bertone2005,Klasen2015, DelPopolo2014}, from small to large scales  \citep{Einasto2001,Bertone2005,Bouchet2004,Kilbinger2015}, one of the largest problems of the model is that the particles that should constitute the DM has never been observed \citep{Klasen2015}.
Moreover, the so called "small scale problems" of the $\Lambda$CDM \citep{DelPopolo2017a} are plaguing the model. Several recipes for overcoming these problems, based on cosmological power spectrum modifications \citep{Zentner2003}, different nature of the DM particles \citep{Colin2000,Goodman2000,Hu2000,Kaplinghat2000,Peebles2000,SommerLarsen2001}, Modified Gravity (MG) theories, such as $f(R)$ \citep{Buchdahl1970,Starobinsky1980}, $f(T)$ \citep{Bengochea2009,Linder2010,Dent2011,Zheng2011}, or MOND \citep{Milgrom1983},
or astrophysical effects \citep{Brooks2013,Onorbe:2015ija,DelPopolo2017a}, have been proposed.
To those issues, we should add that the cosmological constant $\Lambda$ 
suffers from the ``cosmological constant fine tuning problem", and the ``cosmic  coincidence problem" \citep{Astashenok2012,Velten2014,Weinberg1989}. 
 
All those issues motivated the investigations of other explanations, and models to clarify the universe accelerated expansion. These alternative models generate the DE effects through additional matter fields \citep[e.g., quintessence, as in][]{Copeland2006}, or MG models \citep{Horndeski1974,Milgrom1983,Zwiebach1985,Moffat2006,Nojiri2005,Bekenstein2010,DeFelice2010,Linder2010,Milgrom2014,Lovelock1971,Horava2009,Rodriguez2017,Deffayet2010}. 
Disentangling between the plethora of MG models is not an easy task. The solution of the problem, or at least a better understanding of the same, may come from future surveys like:
Euclid\footnote{\url{http://www.euclid-ec.org}}, JDEM\footnote{\url{http://jdem.lbl.gov/}},
SKA\footnote{\url{https://www.skatelescope.org}}, LSST\footnote{\url{https://www.lsst.org}}, or
from new studies of the Cosmic Microwave Background (CMB) \citep{Battye2018a,Battye2018b}.

In addition to the previous issues, we recall that when the {\it Planck} satellite measurements~\citep{Planck:2018vyg} of the CMB anisotropies are compared to low redshift probes, other anomalies appear. We recall that even if the {\it Planck} experiment has measured the CMB power spectra with a very high precision, the cosmological parameters constraints are model-dependent. Of particular importance are the tensions present between the {\it Planck} values for the Hubble constant $H_0$, characterizing the expansion rate of the universe, and local determinations of the Hubble constant, \cite[e.g.][R20]{Riess:2020fzl}. Moreover, the $S_8$ parameter value obtained through the weak lensing experiments~\citep{Joudaki2017,Abbott2018} is often discussed in the literature. 
In this paper, we shall focus on the tension in the Hubble constant $H_0$ between the late time, namely local determinations of the Hubble constant, and early time measurements of the same quantity using the CMB. In a flat $\Lambda$CDM model, the most widely cited prediction from {\it Planck} concerning the Hubble constant is $H_0=67.27 \pm 0.60{\rm \,km\,s^{-1}\,Mpc^{-1}}$ at 68\% confidence level \citep[CL,][]{Planck:2018vyg}. Including the four-point correlation function or trispectrum data, the value becomes $H_0=67.36 \pm 0.54{\rm \,km\,s^{-1}\,Mpc^{-1}}$ at 68\% CL for {\it Planck} 2018 + CMB lensing~\citep{Planck:2018vyg}. We refer to \cite{DiValentino2021} for other measurements and values. In order to measure $H_0$ locally one can measure the distance-redshift relation, by building a distance ladder. The first capability to measure
Cepheids beyond a few Mpc to reach the nearest SNIa hosts was provided by the Hubble Space Telescope (HST), getting the value of $72 \pm 8 \rm km/s/Mpc$ \citep{HST:2000azd} with the Hubble Space Telescope Key Project. This result was later recalibrated 
to obtain $74.3 \pm 2.2 \rm km/s/Mpc$. In 2005, the SH0ES Project (which stands for Supernova, $H_0$, for the Equation of State of Dark Energy) started. The SH0ES Project advanced the previous approach increasing the sample of high quality calibrations of SNIa by Cepheids \citep[R16]{Riess2016} and increasing the number of independent geometric calibrations of Cepheids \citep[R18]{Riess2018}, then applying further improvements (\citeauthor{Riess:2019qba} \citeyear{Riess:2019qba}, R19, see \citeauthor{DiValentino2021} \citeyear{DiValentino2021}). Further improvements come from the ESA Gaia mission Early Data Release 3 (EDR3) of parallax measurements using 75 Milky Way Cepheids with Hubble Space Telescope photometry and EDR3 parallaxes \citep{Riess:2020fzl}, that gives $H_0 =73.2 \pm 1.3 \rm km/s/Mpc$ at 68\% CL \citep[R20 measurement]{Riess:2020fzl}. 
This last measurement is in tension at $4.2 \sigma$ with the Planck value in a CDM scenario. In general, the two different measurements, depending on the datasets considered, give rise to a persisting tension, with 4 $\sigma$ to 6 $\sigma$ disagreement. The discrepancy is so big, that if there is no evidence for systematic errors in the data, it would require a better model which should reduce or make the tensions and anomalies disappear. There are other determinations of $H_0$ widely discussed in \cite{DiValentino2021}. 
A large number of solutions to those discrepancies have been proposed, for which we refer to \cite{DiValentino2021}. One solution, definitely ruled out, is the idea of an under-dense local Universe, which would solve the tension with a sample-variance effect. Another possibility is a departure from isotropy of the expansion of the Universe. Such model have been studied in \cite{LeDelliou:2020kbm}. Anisotropic expansion 
could be detected by estimating the anisotropy in the Hubble constant from SNIa data. Have been proposed scenarios reducing the sound horizon
at recombination. Modifications of the expansion history after recombination, increasing the $H_0$ value and leaving the
sound horizon unaltered are usually dubbed "late time solutions". In the Early Dark Energy (EDE) models (i.e., presence of a significant DE component during the early evolution of the Universe) the Hubble tension can be solved, also reducing at the same time the sound horizon \citep{Karwal:2016vyq,Jiang:2024tll}. Describing all the solutions is almost prohibitive, and, as already reported, a large number has been discussed in \cite{DiValentino2021}. 

The authors of Ref.\citep{Riess:2024vfa} combined all the James Webb Space Telescope (JWST) measurements for each technique, including Cepheids, J-region Asymptotic Giant Branch (JAGB), 
and Tip of the red-giant branch (TRGB) to search for any systematic biases 
and they find $H_0 = 73.4 \pm 2.1$, $72.2 \pm 2.2$, and $72.1 \pm 2.2 ~\rm km/s/Mpc$ for JWST Cepheids, JAGB, and TRGB, respectively. When they combined all the methods (but each SN measurement included only once), they determined  $H_0 = 72.6 \pm 2.0~\rm km/s/Mpc$, in good agreement with $72.8~\rm km/s/Mpc$ that HST Cepheids would yield for the same sample.
\\
The Laser Interferometer Gravitational-Wave Observatory (LIGO) observation of the event GW170817 help to determine a value of $H_0 = 70^{+12}_{-8} ~\rm km/s/Mpc$ by means of the 'standard siren' method which is completely independent of the local distance scale \citep{LIGOScientific:2017adf}.

In this paper, we deal with a Bianchi type I spacetime. \cite{Akarsu2019} claimed that 
an anisotropic correction to $\Lambda$CDM model, obtained by replacing the spatially flat FLRW metric with the Bianchi type-I metric, could reduce the problem. By means of 36 $H(z)$ measurements from Cosmic Chronometer (CC), and Baryonic Acoustic Oscillations (BAO) in galaxy
and Ly-forest distributions, \cite{Akarsu2019} found a value of $H_0 = 70.4 \pm 1.7 \rm km/s/Mpc$ at
68\% CL, in agreement with both the CMB and the R20 values within $2 \sigma$.  
However, a full analysis considering CMB data is still missing. Recently, \cite{Szigeti:2025jxz} showed that a G\"odel inspired slowly rotating universe  resolves this tension with a present angular velocity reaching $\omega_0=2x10^-3 /$Gyr. In the present paper, we dealt with this claim.

We extend the analysis by including the most recent Type Ia supernovae data. Specifically, we use the Pantheon+ compilation \citep{Scolnic:2021amr}, which contains spectroscopically confirmed SNe Ia spanning $0 < z < 2.3$, and represents the most homogeneous and precise dataset to date. For local calibration, we consider the SH0ES measurements, which are based on Cepheid-calibrated SNe Ia and provide strong constraints on the local value of $H_0$. This enables us to test anisotropic models using both early- and late-Universe observables in a consistent framework.

To constrain the model and study the Bayesian inference, we consider the most recent Hubble, SHOE, Pantheon, SCP data relevant to the late Universe ($z \lesssim 2.4$) and then include the baryonic acoustic oscillations
(BAO) and CMB data as well, both of which contain information about the Universe at $z \sim 1100$.

The paper is organized as follows. In Section~\ref{sec:Anisotropic-CDM-model},
we describe the luminosity distance model we used, based on the Ref.~\cite{LeDelliou:2020kbm} model that 
transcribes the Bianchi type-I model into
an apparent almost $\Lambda$CDM model. 
In Section~\ref{sec:ConfrontObs}, we perform a comparative Bayesian analysis using both Markov Chain Monte Carlo and Nested Sampling techniques to study the influence of nuisance parameters and extract robust posterior distributions for the model parameters. In Section~\ref{sec:discussion} we discuss the results and conclude in Section~\ref{sec:conclusions}.

\section{Anisotropic $ \Lambda$CDM Model}
\label{sec:Anisotropic-CDM-model}

Our aim is to propose a model capable of including a level of anisotropies
compatible with observations that is looking very much like a $\Lambda$CDM
model in the past and developing anisotropies into the present. Although this behaviour seems to contradict Wald's theorem \citep{Wald:1983ky}, we built a Bianchi type-I model in \citep{LeDelliou:2020kbm} as an 
anisotropic model for evaluation purpose, with anisotropy only present for a limited time span between the initial time soon before recombination and a short time after present days. Its increase in anisotropy is limited and is not expected to sustain in asymptotic time explored by Wald.
To
do so, \citep{LeDelliou:2020kbm} proposed an expression of a Bianchi type-I model in a form similar
to the 
Friedman--Lemaître--Robertson--Walker
\citep[FLRW,][]{Friedman:1922kd,Lemaitre:1931zz,Robertson:1933zz,Walker:1933} solution, and developed its solution in order to keep
as close as possible to the derivations of the FLRW.

\subsection{Model Setup} 

We follow the same setup as in \citep{LeDelliou:2020kbm} in order to use the model to compute its distance modulus
\begin{figure*}
	\centering
	\subfigure[]{%
	\includegraphics[scale=0.42]{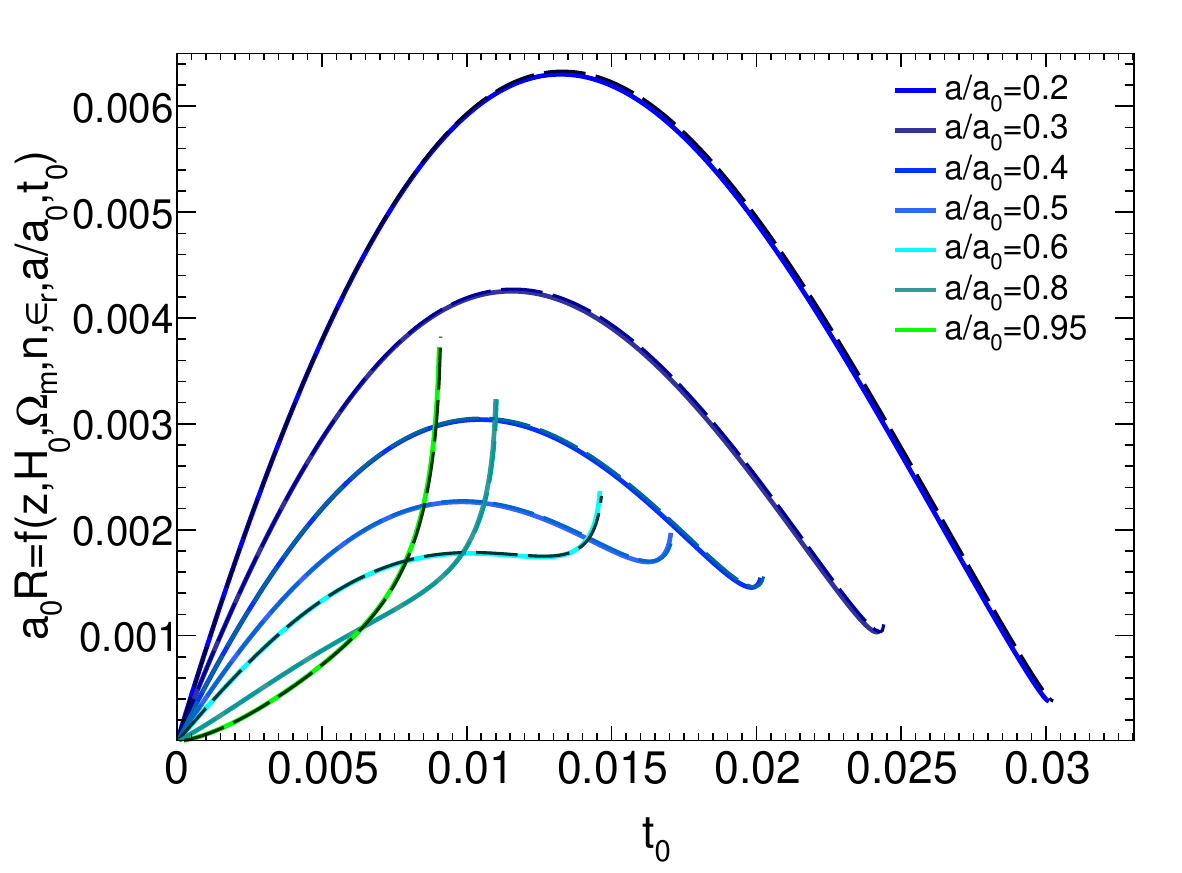}
		\label{fig:integral_a0R}%
	}\quad
	\subfigure[]{%
	\includegraphics[scale=0.42]{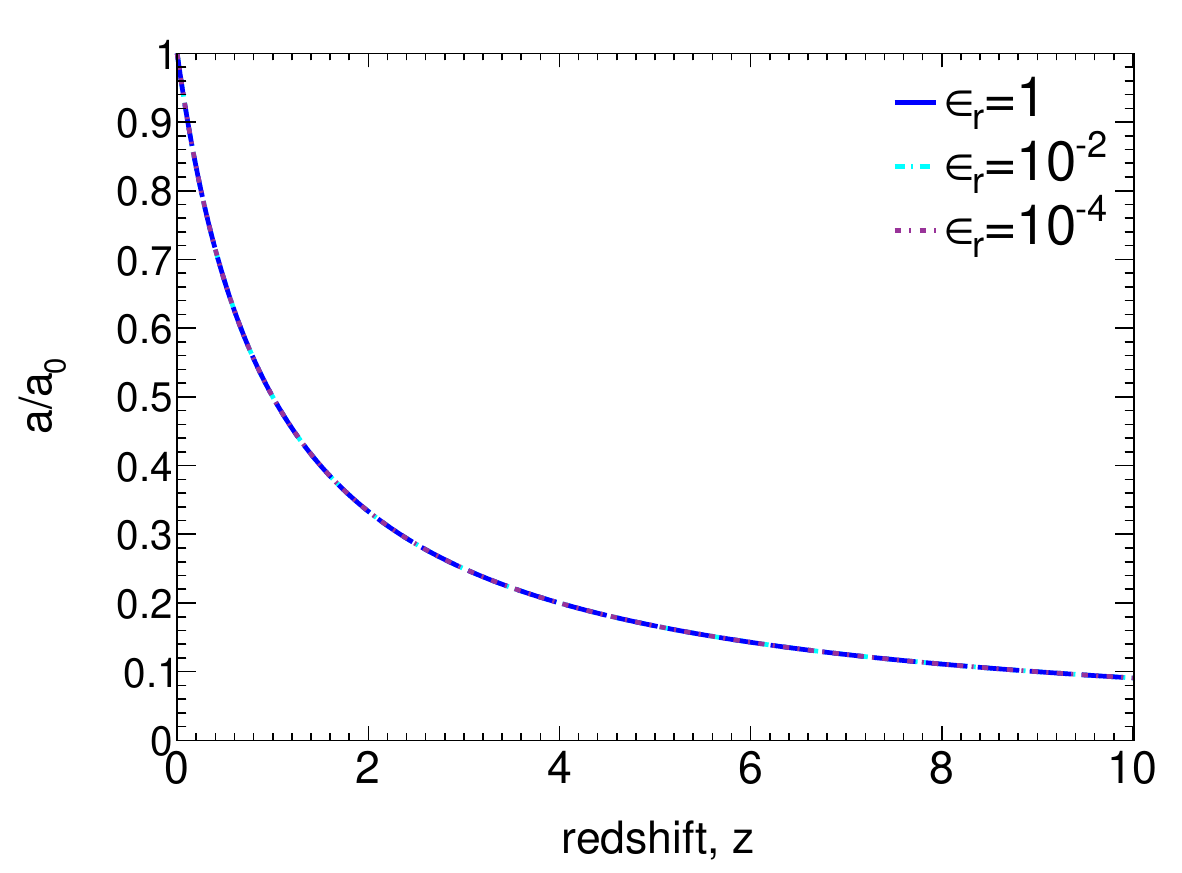}
		\label{fig:aa0_solution}%
	}
	\caption{\justifying 
	(a) Analytical component of the integral from Eq.\eqref{eq:dist_modulus_a0R}, as a function of $t_{0}$ for different ratios, $a/a_{0}$ (see legend).
Solid lines correspond to $H_{0}=67.27~\text{km}\text{s}^{-1}\text{Mpc}^{-1}$ and $\Omega_{m} = 0.21$, while dashed lines represent $\Omega_{m} = 0.25$. The remaining parameters, such as $n$, $\epsilon_{r}$ are arbitrary. All results are presented up to the divergence threshold. 		
	(b) Graph showing the numerical solution to Eq.~\eqref{eq:zOfAsA0}. The choice of $\epsilon_{r}$ has a negligible impact on the evolution of $a/a(0)$: the
	curves corresponding to different $\epsilon_{r}$ values completely overlap.}	
	\label{fig:a0R_and_aa0}
\end{figure*}
\subsection{Distance Modulus Calculation}
\label{sec:Dist_Modulus_Calculation}

The distance modulus is defined as the difference between the apparent (observed) magnitude and absolute magnitude (the magnitude as if the object was at a distance of $a_{0}R$\vspace{-.1cm}
\begin{equation}
	\mu = 5 {\rm{log}} (a_{0}R)-5
	\label{eq:dist_modulus_begining}
\end{equation}\vspace{-.3cm}

\noindent and will depend on the value of the different cosmological parameters and redshift, following \cite[Eq.~(21)]{LeDelliou:2020kbm}, where
\begin{equation}
\begin{split}
R  = \int_{0}^{2\pi}  \int_{t_{e}}^{t_{0}}  \frac{dtd\alpha}{2\pi a \sqrt{(1+\epsilon)^{2} {\rm{cos}}^{2}\alpha + {\rm{sin}}^{2}\alpha}}\\
 =  \int_{t_{e}}^{t_{0}}  \frac{dt}{a}  \int_{0}^{2\pi}  \frac{d\alpha}{2\pi \sqrt{(1+\epsilon)^{2} {\rm{cos}}^{2}\alpha + {\rm{sin}}^{2}\alpha}},
\label{eq:dist_modulus_R}
\end{split}
\end{equation}
where the comoving distance $R$ is integrated over all directions $\alpha$ from the model's largest expansion direction, characterised by the usual scale factor $a$ and the small anisotropy factor $\epsilon$ \cite[see][]{LeDelliou:2020kbm}.
Since $\epsilon$ is small, we approximate the angle average to linear order with
\begin{equation}
\begin{split}
&\int_{0}^{2\pi} \frac{d\alpha}{2\pi \sqrt{(1+\epsilon)^{2} {\rm{cos}}^{2}\alpha + {\rm{sin}}^{2}\alpha}}\\
 &\simeq \int_{0}^{2\pi} \frac{d\alpha}{2\pi \sqrt{1+ 2\epsilon {\rm{cos}}^{2}\alpha }}\\
 &\simeq \int_{0}^{2\pi} \frac{d\alpha}{2\pi} (1 -\epsilon {\rm{cos}}^{2}\alpha) 
 \simeq  1 - \frac{\epsilon}{2}.
\label{eq:dist_modulus_int_sol}
\end{split}
\end{equation}
We obtain $a$ from \cite[Eq.~10, noting their $\Omega_0$ as $\Omega_m$]{LeDelliou:2020kbm}, where $\rho_m$ is a matter energy density, $\Omega_m$ the corresponding energy density parameter, index $0$ represents the present time values, $H_0$ the present Hubble parameter and $\Lambda$ is the cosmological constant, while index $i$ marks some initial time, with\vspace{-.05cm}
\begin{equation}
\begin{split}
\left(\frac{d\rm{ln}a}{dt}\right)^{2} &=\kappa\rho_{m}+\Lambda
 = H_{0}^{2} \left[  \Omega_{m} \left(\frac{a_{0}}{a}\right)^{3} + 1 - \Omega_{m} \right] \\ 
& \Updownarrow \\
t-t_{i} & = \int_{a_{i}}^{a}\frac{da}{H_{0}\sqrt{\Omega_{m} \frac{a_{0}^{3}}{a} + (1-\Omega_{m})a^{2}} }.
\label{eq:approx_a}
\end{split}
\end{equation}
\vspace{.05cm}
From \cite[Eq.~29]{LeDelliou:2020kbm}, for angled averaged quantities, we have
\begin{equation}
		\frac{a_{0}}{a}  = (1+z)\sqrt{\frac{1+(1+\frac{\epsilon}{2})\epsilon}{1+(1+\frac{\epsilon_{0}}{2})\epsilon_{0}}}
		 \simeq (1+z)\sqrt{1+ \epsilon - \epsilon_{0}},
		\label{eq:express_a0a}
\end{equation}
where $z$ expresses a measurable redshift, and, from \cite[Eq.~A5]{LeDelliou:2020kbm}, we link the present anisotropy parameter to its recombination time value, marked by index $r$, with $a_i$ taken at some initial time before recombination and $\Omega_{\Lambda}$ is the energy density parameter for $\Lambda$,
\begin{equation}
 	\epsilon_{0} = \frac{\epsilon_{r}}{1-\frac{\sqrt{\Omega_{m}\left(\frac{a_{r}}{a_{0}}\right)^{-3} + \Omega_{\Lambda}} - 1}{\sqrt{\Omega_{m}\left(\frac{a_{i}}{a_{0}}\right)^{-3} + \Omega_{\Lambda}} - 1}} .
	\label{eq:epsilon0}
\end{equation}

Solution of Eq.~\eqref{eq:approx_a} to growing scale evolution from initial $t=0$ and $a=0$ at Big Bang yields (see Appendix~\ref{sec:AppFriedmannSol} for details)

\begin{equation}
 a=a_{0}e^{-\sqrt{\Omega_{\Lambda}}H_{0}\left(t-t_0\right)}\left[\frac{e^{3\sqrt{\Omega_{\Lambda}}H_{0}t}-1}{e^{3\sqrt{\Omega_{\Lambda}}H_{0}t_0}-1}\right]^{\frac{2}{3}}.\label{eq:BigBang_solution_t0}
\end{equation}

Setting present time $a_{0} = 1$ and recombination $a_{r} = 10^{-3}$ one can solve for $t_{0}$ and $t_{r}$. As from \cite[Eq.~16]{LeDelliou:2020kbm}, the anisotropy $\epsilon$ remains small, 
we approximate the physical distance with 
\begin{equation}
a_{0}R = \int_{t_{e}}^{t_{0}} \frac{a_{0} dt}{a} \left(1-\frac{\epsilon}{2}\right)
\simeq \left(1-\frac{\epsilon}{2}\right)  \int_{t_{e}}^{t_{0}}  \frac{a_{0}}{a}dt.
\label{eq:dist_modulus_a0R}
\end{equation}	
The analytical part of the distance integral in Eq.\eqref{eq:dist_modulus_a0R} is illustrated in Fig.\ref{fig:integral_a0R}, where it is plotted as a function of $t_0$ for several fixed values of the scale factor ratio $a/a_0$. The behavior of this term is shown for two values of $\Omega_m$, with all other parameters held arbitrary.

With the 
solution \eqref{eq:BigBang_solution_t0}, the distance modulus reads (see Appendix \ref{sec:LumDistInt})

\begin{align}
\mu= & 5\log\left\{ \left(1-\frac{1}{2}\epsilon\right)\frac{e^{-H_{0}\sqrt{1-\Omega_{m}}t_{0}}}{H_{0}\sqrt{1-\Omega_{m}}}\left[\left(e^{3H_{0}\sqrt{1-\Omega_{m}}t_{0}}-1\right)\right.\right.\nonumber \\
 & \hspace{2.5cm}\times~{}_{2}F_{1}\left(\frac{1}{3};\frac{2}{3};\frac{4}{3};1-e^{3H_{0}\sqrt{1-\Omega_{m}}t_{0}}\right)\nonumber \\
 & \hspace{1cm}-\left(e^{3H_{0}\sqrt{1-\Omega_{m}}t_{0}}-1\right)^{\frac{2}{3}}\left(e^{3H_{0}\sqrt{1-\Omega_{m}}t}-1\right)^{\frac{1}{3}}\nonumber\\
 & \left.\vphantom{\frac{e^{-H_{0}\sqrt{1-\Omega_{m}}t_{0}}}{H_{0}\sqrt{1-\Omega_{m}}}}\left.\hspace{2.5cm}\times~{}_{2}F_{1}\left(\frac{1}{3};\frac{2}{3};\frac{4}{3};1-e^{3H_{0}\sqrt{1-\Omega_{m}}t}\right)\right]\right\} -5,
\end{align}
and since from Appendix \ref{sec:AnisoParam}, we have $\frac{\epsilon}{\epsilon_{r}}\simeq10^{-\frac{9}{2}}\left(\frac{a}{a_{0}}\right)^{-\frac{3}{2}}$, the final luminosity distance reads
\begin{align}
\mu= & 5\log\left\{ \left(1-\frac{10^{-\frac{9}{2}}}{2}e^{\frac{3}{2}\sqrt{\Omega_{\Lambda}}H_{0}\left(t-t_{0}\right)}\left[\frac{e^{3\sqrt{\Omega_{\Lambda}}H_{0}t_{0}}-1}{e^{3\sqrt{\Omega_{\Lambda}}H_{0}t}-1}\right]\epsilon_{r}\right)\right.\nonumber \\
 & \hspace{1cm}\times\frac{e^{-H_{0}\sqrt{1-\Omega_{m}}t_{0}}}{H_{0}\sqrt{1-\Omega_{m}}}\nonumber \\
 & \hspace{1.5cm}\times\left[\vphantom{\left(e^{3H_{0}\sqrt{1-\Omega_{m}}t_{0}}-1\right)^{\frac{2}{3}}}\left(e^{3H_{0}\sqrt{1-\Omega_{m}}t_{0}}-1\right)\right.\nonumber \\
 & \hspace{2cm}\times~{}_{2}F_{1}\left(\frac{1}{3};\frac{2}{3};\frac{4}{3};1-e^{3H_{0}\sqrt{1-\Omega_{m}}t_{0}}\right)\nonumber \\
 & \hspace{1cm}-\left(e^{3H_{0}\sqrt{1-\Omega_{m}}t_{0}}-1\right)^{\frac{2}{3}}\left(e^{3H_{0}\sqrt{1-\Omega_{m}}t}-1\right)^{\frac{1}{3}}\nonumber \\
 & \hspace{1.5cm}\left.\vphantom{\frac{e^{-H_{0}\sqrt{1-\Omega_{m}}t_{0}}}{H_{0}\sqrt{1-\Omega_{m}}}}\left.\times~{}_{2}F_{1}\left(\frac{1}{3};\frac{2}{3};\frac{4}{3};1-e^{3H_{0}\sqrt{1-\Omega_{m}}t}\right)\right]\right\} -5,\label{eq:muOfT}
\end{align}

where ${}_{2}F_{1}$ is the hypergeometric function.

Inverting Eq.~\eqref{eq:BigBang_solution_t0} as in Appendix~\ref{sec:AppFriedmannSol} the time dependence follows the scale as

\begin{align}
t= & \frac{2}{3\sqrt{\Omega_{\Lambda}}H_{0}}\ln\left\{ \vphantom{\sqrt{\frac{\left(e^{3At_{0}}-1\right)^{2}}{4}}}\frac{\left(e^{3\sqrt{\Omega_{\Lambda}}H_{0}t_{0}}-1\right)}{2}\left(\frac{a/a_{0}}{e^{\sqrt{\Omega_{\Lambda}}H_{0}t_{0}}}\right)^{\frac{3}{2}}\right.\nonumber\\
 & \left.+\sqrt{1-\frac{\left(e^{3\sqrt{\Omega_{\Lambda}}H_{0}t_{0}}-1\right)^{2}}{4}\left(\frac{a/a_{0}}{e^{\sqrt{\Omega_{\Lambda}}H_{0}t_{0}}}\right)^{3}}\right\} ,\label{eq:tOfAsA0}
\end{align}
with $t_{0}$ solution of
\begin{align}
\hspace{-.5cm}t_{0}= & \frac{2}{3\sqrt{\Omega_{\Lambda}}H_{0}}\ln\left\{ \sinh\left(\frac{3}{2}\sqrt{\Omega_{\Lambda}}H_{0}t_{0}\right)+\sqrt{1-\sinh^{2}\left(\frac{3}{2}\sqrt{\Omega_{\Lambda}}H_{0}t_{0}\right)}\right\} .\label{eq:solT0}
\end{align}
The redshift time dependent is then given by inverting Eq.~\eqref{eq:express_a0a}, with the approximation of Appendix~\ref{sec:AnisoParam}
\begin{align}
z= & \frac{1}{\sqrt{\left(\frac{a}{a_{0}}\right)^{2}\left(1-10^{-\frac{9}{2}}\epsilon_{r}\right)+10^{-\frac{9}{2}}\epsilon_{r}\left(\frac{a}{a_{0}}\right)^{\frac{1}{2}}}}-1,
\label{eq:zOfAsA0}
\end{align}
which allows us, combining Eqs.~\eqref{eq:muOfT} and \eqref{eq:tOfAsA0} with \eqref{eq:zOfAsA0}, to obtain the redshift dependence of the distance modulus $\mu\left(z\right)$, solving the value of the present cosmic time $t_0$ from Eq.~\eqref{eq:solT0}.

The numerical inversion of Eq.\eqref{eq:zOfAsA0} is illustrated in Fig.\ref{fig:aa0_solution}, where we show the evolution of $a/a(0)$ as a function of redshift for various values of $\epsilon_r$. As evident, the choice of $\epsilon_r$ has a negligible effect on the result, with all curves overlapping.

\begin{table*}
	\centering
		\begin{tabular}{l|l|l}
			\hline
			\hline
			method & parameters &prior type\\
			\hline                
			\texttt{bilby, PyMultiNest, nessai} 
			&$H_{0}^{\rm{anis}}$ [km s$^{-1}$ Mpc$^{-1}$]  &$\mathcal{U}(55;85)$\\
            &$\Omega_{m}^{\rm{anis}}$   &$\mathcal{U}(0.1;0.5)$\\
            &$\epsilon_{r}$ &$\mathcal{U}(0;1\times 10^{-4})$\\ 
            &$n$ &$\mathcal{U}(0.1;4.1)$\\
			\hline  
			\texttt{bilby, PyMultiNest, nessai} 
			&$H_{0}^{\rm{anis}}$ [km s$^{-1}$ Mpc$^{-1}$] &$\mathcal{G}(72;10)$\\
			&$\Omega_{m}^{\rm{anis}}$  &$\mathcal{G}(0.25;0.06)$\\
            &$\epsilon_{r}$  &$\mathcal{G}(3\times 10^{-4};3\times 10^{-2})$\\  
            &$n$  &$\mathcal{G}(1.1;2)$\\           
			\hline
			\hline
		\end{tabular}
		\caption{Priors for the Bianchi type-I  model of Ref.~\citep{LeDelliou:2020kbm}. Most probable intervals of the model parameters (90\% confidence level) constrained by our analysis for the two priors types: uniform ($\mathcal{U}(\text{min}, \text{max})$) and Gaussian ($\mathcal{G}(\mu,\sigma)$) distributions.}
	\label{tab:priors_intervals}		
\end{table*}

\section{Confrontation with observations}
\label{sec:ConfrontObs}
\subsection{The Study of the Nuisance Parameters Used with Bayesian inference}
\label{sec:Study_NuisanceParametersUsed}

Bayesian inference provides a useful approach to parameter estimation and model selection. In this section we describe some of the key concepts of Bayesian inference and how it can be used in the task of fitting a model.  

For the case of $H_{0}$ modeling, we use Markov Chain Monte Carlo (MCMC)
analysis in the form of an ensemble sampler \citep{10.1111/j.2517-6161.1995.tb02042.x, 10.1214/aos/1056562461} and Nested Sampling (NS) analysis in the form of nested elliptical sampling\citep{Skilling:10.1063/1.1835238}. Without going into these methods  
details, 
they are employed in our paper for the  
sake of comparative analysis, in order  
to recover the posteriors distributions of the model parameters.
This was done by 
parameter estimation packages such as \texttt{bilby} \citep{Ashton:2018jfp,Smith:2019ucc}, \texttt{PyMultiNest} \citep{Buchner:2014nha} and \texttt{nessai} \citep{Williams:2023ppp}. 
This allowed
us to not only have a powerful tool for simulating the Bayesian samples for our model, but to have a full suite ready to quickly analyze and fit datasets producing high level statistical information. 
By having comprehensive statistical information that goes beyond assigning a likelihood value to a set of parameters, we also obtain converged probability distributions of the parameters and their correlations along with a measure of their Bayesian evidence that can be used for model comparisons.

\subsubsection{Bayesian inference}
\label{sec:BayesianInference}

We define a dataset $D$ as the combination of observations $y$ and their associated errors $\sigma$. A model $\mathcal{M}$ is defined as
some function that can be applied to a set of parameters $\theta$
to obtain some model predictions $y_{\mathcal{M}}$ that can, in turn,  be compared with the dataset. Then, we can define the probability that a dataset has been obtained given a model and a set or parameters as:
\begin{equation}
P(D\vert \theta, \mathcal{M}) \equiv \mathcal{L},
\label{eq:BaysParam}
\end{equation}
where $\mathcal{L}$ is called the likelihood. Likewise, we can also define
the {\it a priori} probability, or belief, of the parameters that we
have used:
\begin{equation}
P( \theta \vert \mathcal{M}) \equiv \pi,
\label{eq:BaysPrior}
\end{equation}
where $\pi$ is denoted the prior. For example, for anisotropic models, the parameters that most determine this division are $H_{0}$, $n$ and $\varepsilon$. If we have any a priori information suggesting a certain distribution for $H_{0}$, $n$ and $\varepsilon$, we can include this information as {\it a priori}. If not, the prior is typically set to be flat. Note however that this does not mean that no prior assumptions were included: a flat prior in logarithmic space will for example have a different impact than
a flat prior in linear space.

By integrating over the whole parameter space $\Theta$, we can obtain the probability of obtaining a dataset given a model:
\begin{equation}
P( D \ \vert \mathcal{M}) \equiv \mathcal{Z} = \int_{\Theta} P(D\vert \theta, \mathcal{M}) P(\theta \vert \mathcal{M}) d\theta,
\label{eq:BaysProb}
\end{equation}
where $\mathcal{Z}$ is the Bayesian evidence of the model, simplifying the
notation:
\begin{equation}
\mathcal{Z} \equiv p(D\vert \mathcal{M})=\int_{\Theta} \mathcal{L}(\theta) \pi(\theta) d\theta
\label{eq:BayesEvidence}
\end{equation}
The evidence can be understood as a weighted average of the likelihood $\mathcal{L}$ over the whole parameter space with the prior $\pi$ as a weight function. Thus, only the regions where the product of both the likelihood and the prior is high will mainly contribute to the value of the evidence. By integrating over the whole parameter space with a normalized prior,
we are implicitly dividing by the volume of the space, so models with more parameters that do not improve the likelihood will obtain a lower evidence.

By applying Bayes’ theorem, we can find the probability of a model given a dataset
\begin{equation}
P( \mathcal{M} \vert D) = \frac{P(D\vert\mathcal{M}) P(\mathcal{M})}{P(D)},
\label{eq:BayesTheor}
\end{equation}
where $P(\mathcal{M})$ and $P(D)$ are the {\it a priori} probability of, or belief in,  
the model and the data, respectively. Here, $P(D)$ denotes the probability of measuring the dataset $D$ from the underlying physics and 
observation method. $P(\mathcal{M})$ encapsulates our prior belief in the model M that is not already reflected in the priors of the parameters. For example, if we know that $H_{0}$ could be fit from two different Bianchi type-I models with a ratio of 1 to 4, we can incorporate this information into this term.

\subsubsection{Bayes factor}
\label{sec:BayesFactor}

Denoting two models by $a$ and $b$, we can take the ratio of their probabilities to obtain:
\begin{equation}
R= \frac{P(\mathcal{M}_{a}\vert D)}{P(\mathcal{M}_{b}\vert D)}=\frac{P(D\vert \mathcal{M}_{a})P(\mathcal{M}_{a})}{P(D\vert \mathcal{M}_{b})P(\mathcal{M}_{b})} = \frac{\mathcal{Z}_{a}P(\mathcal{M}_{a})}{\mathcal{Z}_{b}P(\mathcal{M}_{b})},
\label{eq:BayesFactor}
\end{equation}
where $\mathcal{Z}_{a}$ and $\mathcal{Z}_{b}$ are defined by Eq.\eqref{eq:BayesEvidence},
$P(\mathcal{M}_{a})/P(\mathcal{M}_{b})$ is the prior probability ratio between the two models and $P(\mathcal{M}_{a}\vert D)/P(\mathcal{M}_{b}\vert D)$ is the posterior probability ratio of the two models, given the data set $D$.

Assuming that both models have the same {\it a priori} probability, we are left with the ratio of evidences as a way of comparing two models, so that if $R \textgreater 1$ model $a$ is more likely and, conversely
, if $R \textless 1$, model $b$ is more likely, given the data. As with all criteria for model selection, the actual amount of (relative) confidence in a given model that is linked to a given numerical value of $R$ remains subjective. Taking Ref.~\citep{doi:10.1080/01621459.1995.10476572} as a guideline, we give the following judgements as a function of 
the value of $R: 1- 3.2$, `not worth more than a bare mention'; $3.2-10$, `substantial'; $10-100$, `strong' and $\textgreater 100$, `decisive'.

By inverting Eq.\eqref{eq:BaysPrior} and the other conditions, Eqs.~\eqref{eq:BaysParam} and \eqref{eq:BayesEvidence}, we can find the probability of a point in parameter space given a dataset and a model
\begin{equation}
P(\theta \vert D, \mathcal{M})= \frac{P(D\vert \theta, \mathcal{M})P(\theta \vert \mathcal{M})}{P(D\vert\mathcal{M})},
\label{eq:BayesFactor_invert}
\end{equation}
which we call the posterior distribution of $\theta : \mathcal{P(\theta)}$. Simplifying the notation, we obtain:
\begin{equation}
\mathcal{P(\theta)} = \frac{\mathcal{L}(\theta) \pi(\theta)}{\mathcal{Z}}.
\label{eq:PosteriorDistribution}
\end{equation}
For the purpose of parameter estimation, it is not necessary to compute the evidence.

Assuming that the data follow a Gaussian distribution, we can define the log-likelihood\footnote{The reader should not confuse the Likelihood $(\mathcal{L})$ with the log-likelihood $(L)$.} associated between a dataset and the model predictions as:
\begin{equation}
L=-\frac{1}{2} \sum_{i}^{n} \left[ {\rm{ln}}(2\pi \sigma_{i}^{2}) + \left( \frac{y_{i} - y_{\mathcal{M},i}}{\sigma_{i}} \right)^{2} \right]
\label{eq:log-likelihood}
\end{equation}
The first term is a constant for every log-likelihood, as it does not depend on the model, only on the errors of the data. It can be proven that this term appears as a constant multiplicative factor upon the calculation of the evidence, so it divides out when comparing the evidence of two models.

\subsubsection{MCMC and Nested Sampling}
\label{sec:MCMC_NS}

MCMC and Nested Sampling are two different approaches to analyse a model, using Bayesian inference, given its likelihood. MCMC can sample from a given function, which in this case 
is the likelihood of the model, reconstructing thence 
the shape of the likelihood and the posterior probabilities of any of its parameters. Nested sampling works by sampling from the likelihood function in a monotonically increasing way. By assigning a statistical weight to each of the samples, the shape of the likelihood can be reconstructed, and the evidence of the model can be calculated.

In contrast to MCMC analysis, nested sampling 
is aimed towards 
calculation of the evidence. The basic idea is that, by creating a sequence of points with increasing 
likelihood and by assigning a weight to each of them depending on their position in the sequence, the evidence can be calculated. As the value of the likelihood increases in the sequence, the probability of drawing a new point with a better value decreases. \texttt{MultiNest} avoids this problem by using a set of points as ’live points’ and using them to define ellipses from where to drawn the new points \citep{Feroz:2008xx}. This way, the sampling is much more efficient and the sequence can grow faster.

As the set of `live points' define the sampled volume, it can happen that this volume can be separated into isolated sub-volumes. In these cases, each of the sub-volumes is associated with one maximum of the likelihood function, and Nested Sampling, and \texttt{MultiNest} by extension, can detect and handle these separated maxima with ease.

Even though the length of the sequence can be made arbitrarily large
, there is a sequence for which 
the last points contribute very little to the calculation of the evidence. \texttt{MultiNest}, by default, is set to stop the calculation of the evidence when the best point of the sequence increases the value of the log evidence by less than some tolerance.

Each of the points in the sequence is assigned a weight, and thus, can be used to create the histograms of the posterior probabilities of the parameters.

We have used the Bayesian analysis \texttt{PyMultiNest} software \citep{Buchner:2014nha},
which is one of the state-of-the-art nested samplers, to estimate the modelling parameters and associated errors. 
\texttt{MultiNest} \citep{Feroz:2008xx} is a Bayesian inference method, based on the ideas of nested sampling \citep{Skilling:10.1063/1.1835238}. \texttt{MultiNest} uses a form of rejection sampling: it repeatedly draws samples from the prior distribution, under the restriction that the likelihood values are above a certain threshold that increases during the run of the analysis.

For the $H_{0}$, $\Omega_{m}$, $n$-factor, and $\epsilon_{r}$  model parameters, we generate $n_{\rm{MC}}$ Monte Carlo points from their priors. We used $n_{\rm{MC}} = 100 000$ to ensure a dense sampling.
The number of live points for \texttt{PyMultiNest} was 
chosen in such a way that it is sufficient 
to explore the four dimensional parameter space defined by the $H_{0}$, $\Omega_{m}$, $n$-factor, and $\epsilon_{r}$. 

We have set $\texttt{nlive} = 5000$. For comparison, we also performed parameter estimation with the standard Bayesian framework 
with three different samplers: \texttt{bilby}, \texttt{nessai} and \texttt{PyMultiNest}.

\texttt{MultiNest} provides the natural logarithm of the Bayesian evidence, thus one can easily compute $\rm{ln}(\mathcal{Z}_{1}/\mathcal{Z}_{2})$, i.e. the natural logarithm of the Bayes factor between any two tested models. Given
that we consider the prior probabilities of the models to be equal
$p(M_{1})/p(M_{2})$, the Bayes factor is a direct indication of whether a model has a higher probability to be correct than another, given a
data set.

In our Bayesian analysis, we consider both uniform ($\mathcal{U}$) and Gaussian ($\mathcal{G}$) priors for key model parameters. The uniform priors are chosen to reflect a state of minimal prior knowledge, ensuring that the inference is primarily data-driven. These priors help us to remain agnostic about the parameter distribution within a physically or observationally motivated range. 
On the other hand, Gaussian priors are applied where prior measurements or theoretical constraints already exist. Usage of the Gaussian priors allows us to incorporate external knowledge while quantifying how strongly the data update or support these assumptions. This dual-prior approach provides a more complete and systematic way to evaluate the evidence for various cosmological scenarios of the examined model. 
Our prior assumptions are summarized in Table~\ref{tab:priors_intervals}.

\begin{figure*}
\subfigure[]{\includegraphics[scale=0.44]{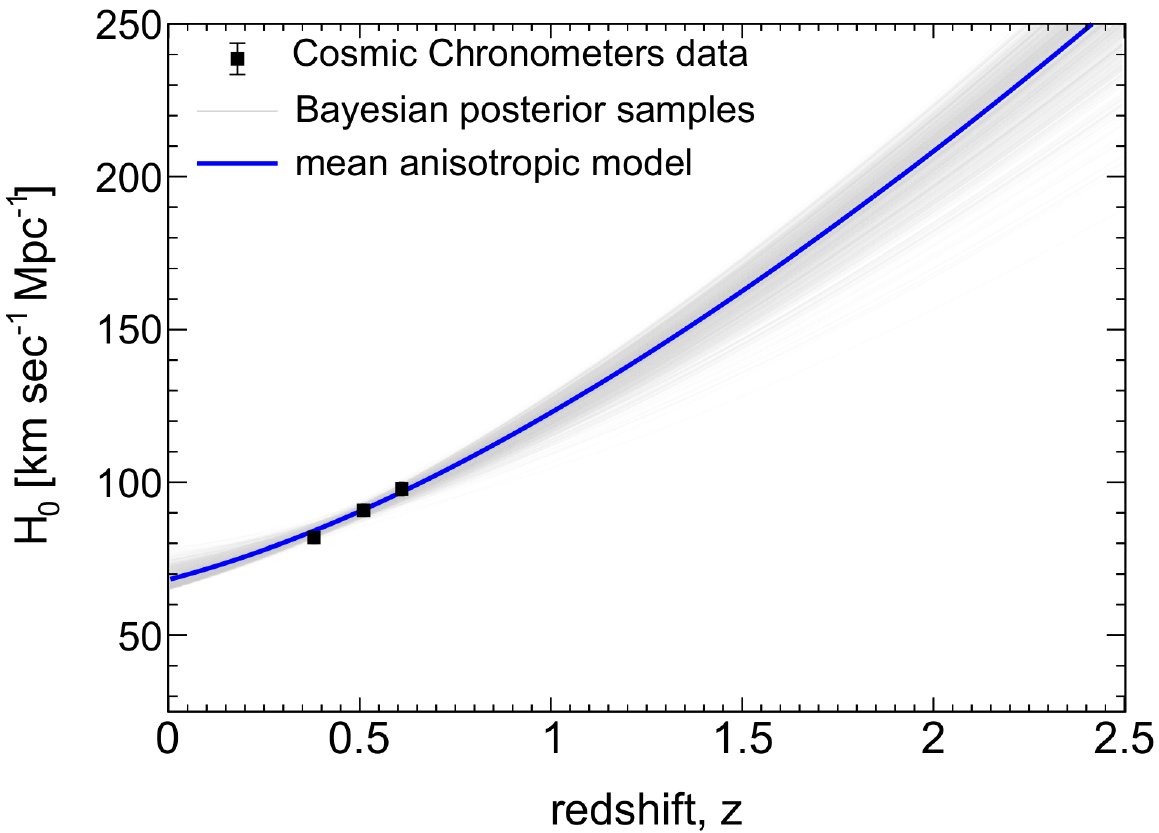}}	
\subfigure[]{\includegraphics[scale=0.44]{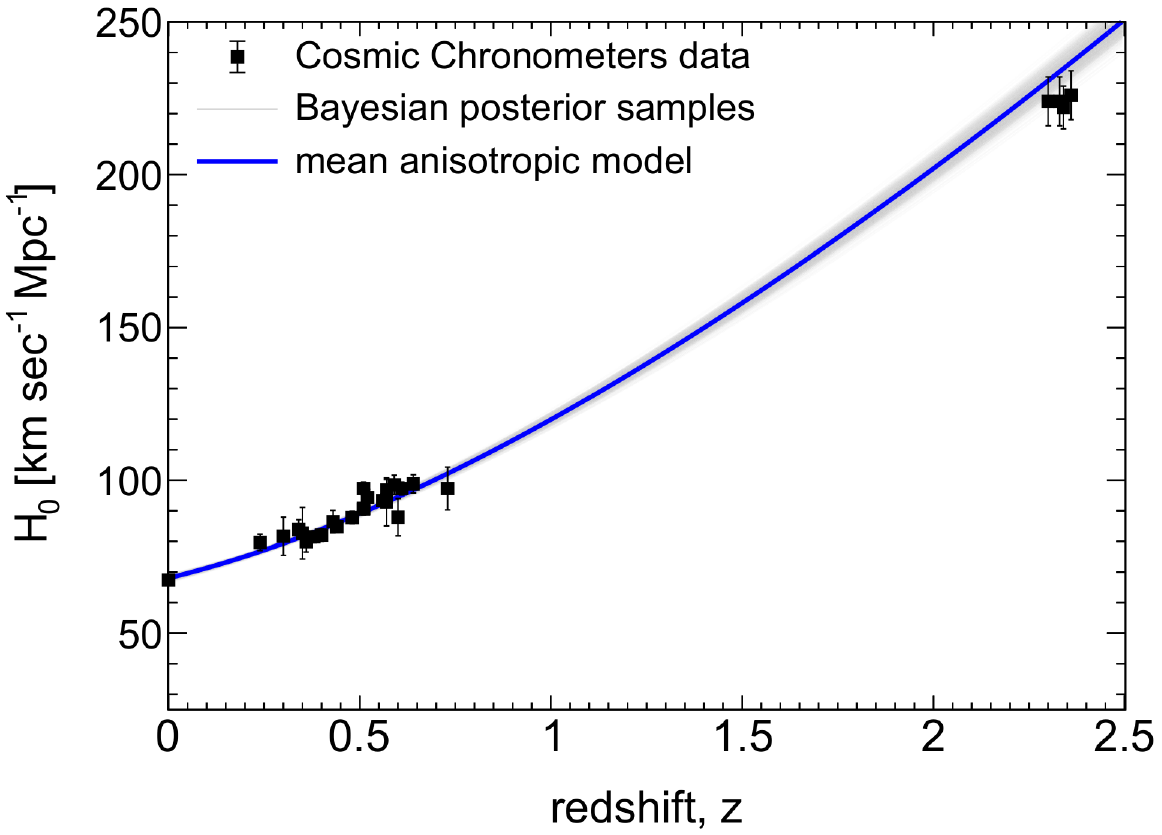}}
\subfigure[]{\includegraphics[scale=0.44]{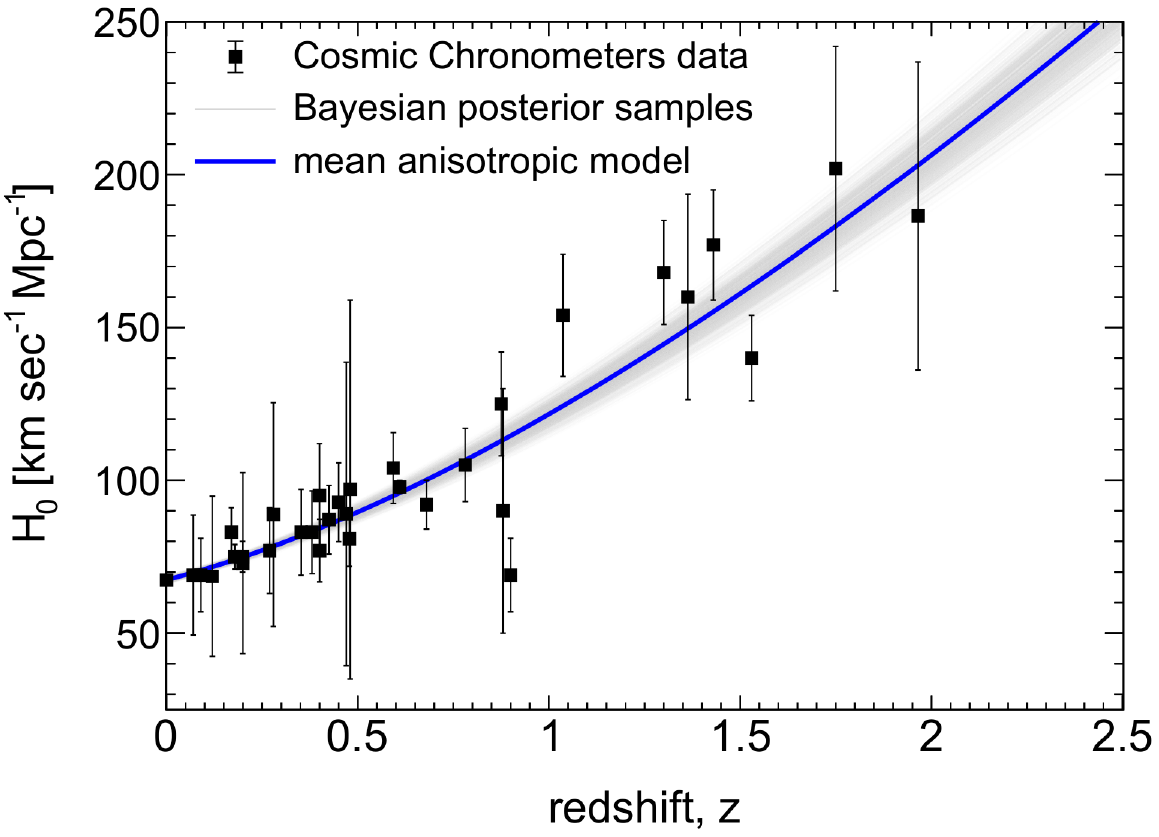}}	
\subfigure[]{\includegraphics[scale=0.44]{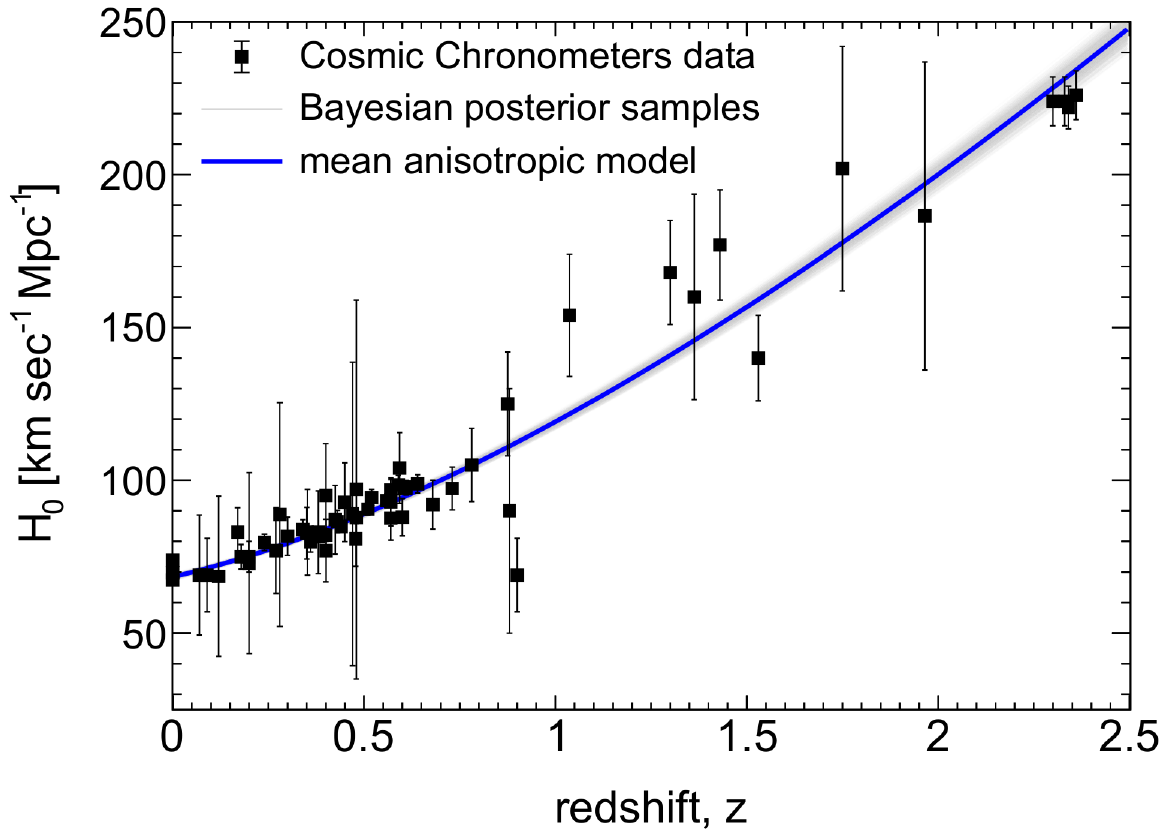}}
\caption{\justifying 
Hubble diagram. $H_{0}$ of observational CC and BAO data vs. $z$. 
(a). The 3 dots represent the galaxy distribution measurements at $z=0.38; 0.51; 0.61$, see Sect.\ref{sec:DataUsed} for mode details; 
(b) Hubble parameter observation measured with BAO method; 
(c) CC observations measured with DA method; 
(d) combined sample CC and BAO datasets fitted with \texttt{bilby} with uniform prior PDFs for $\epsilon_{r}$ and $n$.
}
\label{fig:result_fit_dist_CC}		
\end{figure*}

\subsection{Application to Anisotropic model}
\label{sec:app_aModel}

\label{sec:DataUsed}

We consider a compilation of 68 $H(z)$ measurements as shown in Table \ref{tab:zi_H0obs_our}, viz., the first 3 measurements obtained using CMB \citep{Planck:2018vyg}, 
TRGB 
\citep{Freedman:2020dne}, and Cepheids \cite[anchored to NGC 4258 + MW,][]{Riess:2021jrx} methods.
\textbf{33} measurements have been obtained using Differential Age (DA) method \citep{Jimenez:2001gg}, which is a key approach used in the Cosmic Chronometer (CC) observations \citep{Moresco:2012jh} to estimate the Hubble parameter. 
\textbf{23} measurements have been obtained by the BAO method, where three correlated measurements (at $z=0.38$, $z=0.51$ and $z=0.61$) came from the BAO signal in galaxy distribution \citep{BOSS:2016wmc}, and the last two measurements (at $z=2.34$, $z=2.36$) were determined from the BAO signal in the Ly-$\alpha$ forest distribution alone or cross-correlated with quasistellar objects (QSOs) \citep{BOSS:2014hwf,BOSS:2013igd}. The chi-squared function for the \textbf{59} $H(z)$ measurements, denoted by $\chi^{2}_{CC+\rm{Ly}\alpha}$, is
\begin{equation}
\chi^{2}_{CC+\rm{Ly}\alpha} = \sum_{i=1}^{59} \frac{[H_{\rm{obs}}(z_{i})- H_{\rm{th}}(z_{i})]^{2} }{\sigma^{2}_{H_{\rm{obs}}(z_{i})}},
\label{eq:chi2_CC_Lya}
\end{equation}
where $H_{\rm{obs}}(z_{i})$ is the observed value of the Hubble parameter at the redshift, $z$, with the standard deviation $\sigma^{2}_{H_{\rm{obs}}(z_{i})}$ as given in the Table \ref{tab:zi_H0obs_our} and $H_{\rm{th}}(z_{i})$ is the theoretical value obtained from the cosmological model under consideration.

On the other hand, the covariance matrix related to the three measurements from galaxy distribution \citep{BOSS:2016wmc} reads
\begin{equation}
C=
\begin{bmatrix}
	3.65 &1.78 &0.93\\
	1.78 &3.65 &2.20\\
	0.93 &2.20 &4.45	
\end{bmatrix}
.
\label{eq:CovMatrix_CC}	
\end{equation}
The $\chi^{2}$-function for the three galaxy distribution measurements is
\begin{equation}
	\chi^{2}_{\rm{Galaxy}} = M^{T}C^{-1}M,	
	\label{eq:chi2_Galaxy}
\end{equation}
where
\begin{equation}
	M=
	\begin{bmatrix}
		H_{\rm{obs}}(0.38)-\mathcal{H}(0.38)\\
		H_{\rm{obs}}(0.51)-\mathcal{H}(0.51)\\
		H_{\rm{obs}}(0.61)-\mathcal{H}(0.61)
	\end{bmatrix},
	\label{eq:MMatrix}	
\end{equation}
where $\mathcal{H}$ defined by Eq.~(47) from \cite{LeDelliou:2020kbm}
. 
Henceforth, the combined $\chi^{2}$-function for $H(z)$ measurements, denoted by $\chi^{2}_{\rm{H}}$, is
\begin{equation}
	\chi^{2}_{\rm{H}} = \chi^{2}_{CC+\rm{Ly}\alpha} - \chi^{2}_{\rm{Galaxy}}.
	\label{eq:chi2_H}
\end{equation}

\begin{figure*}
\begin{minipage}[t]{1\paperwidth}%
\hspace{-1.2cm}\includegraphics[scale=0.5]{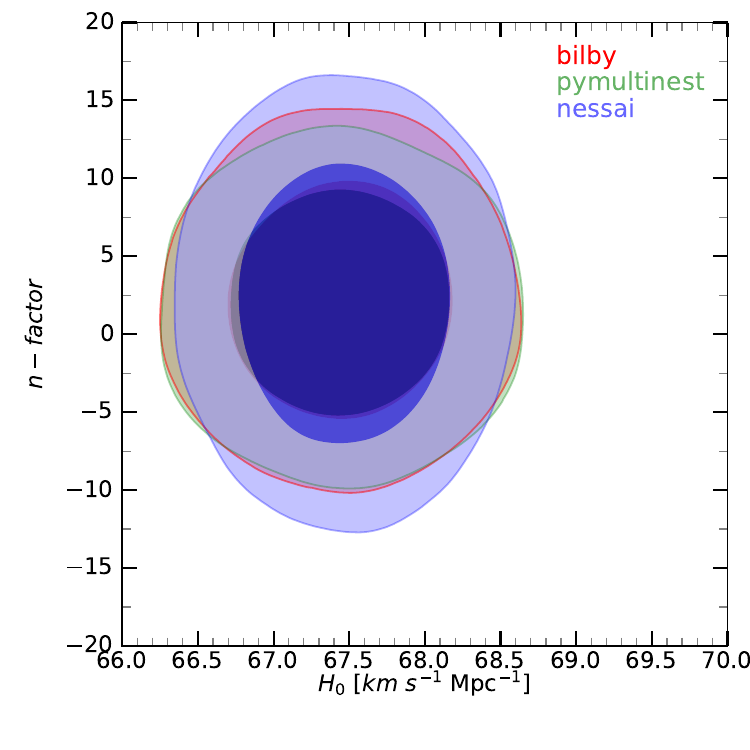}
\includegraphics[scale=0.5]{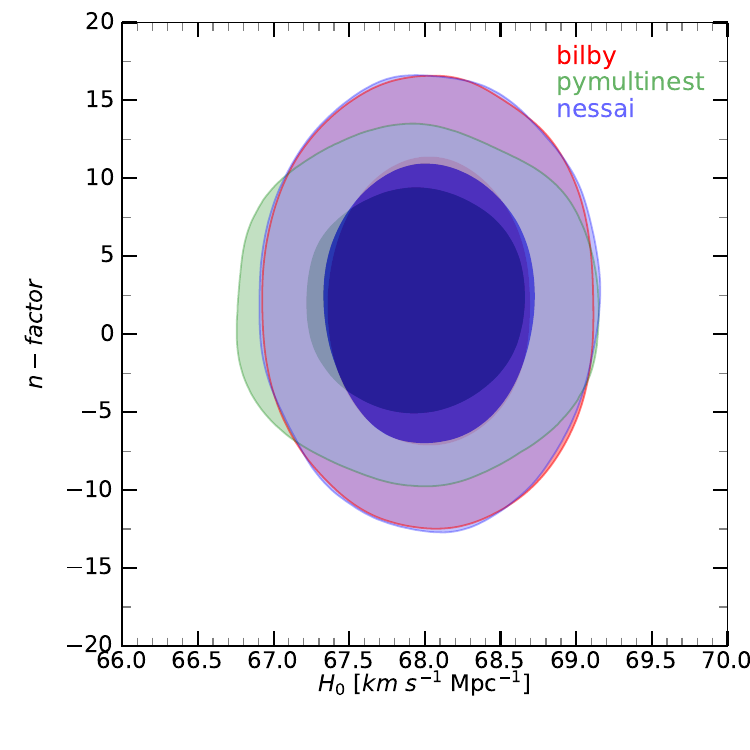}
\includegraphics[scale=0.5]{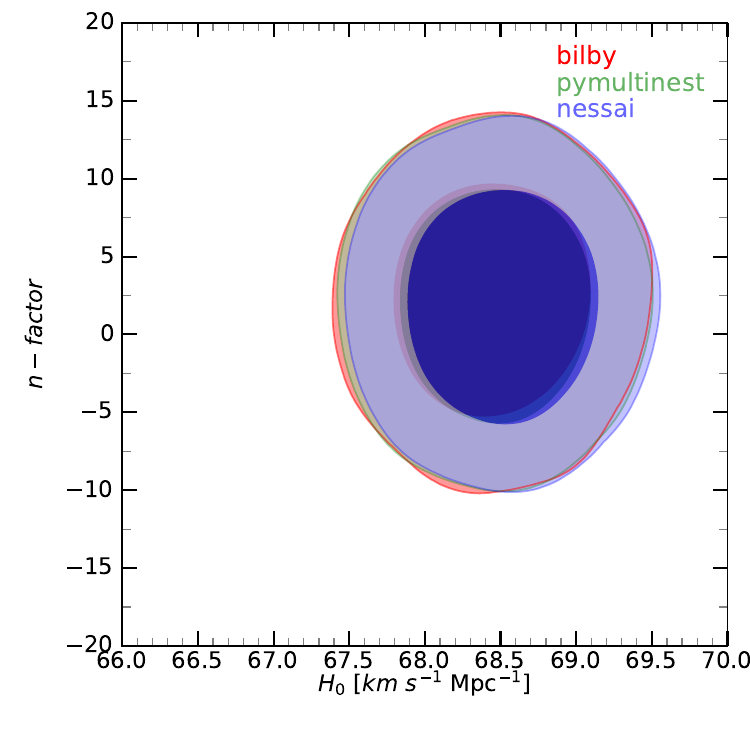}
\end{minipage}

\begin{minipage}[t]{1\paperwidth}%
\hspace{-1.2cm}\includegraphics[scale=0.5]{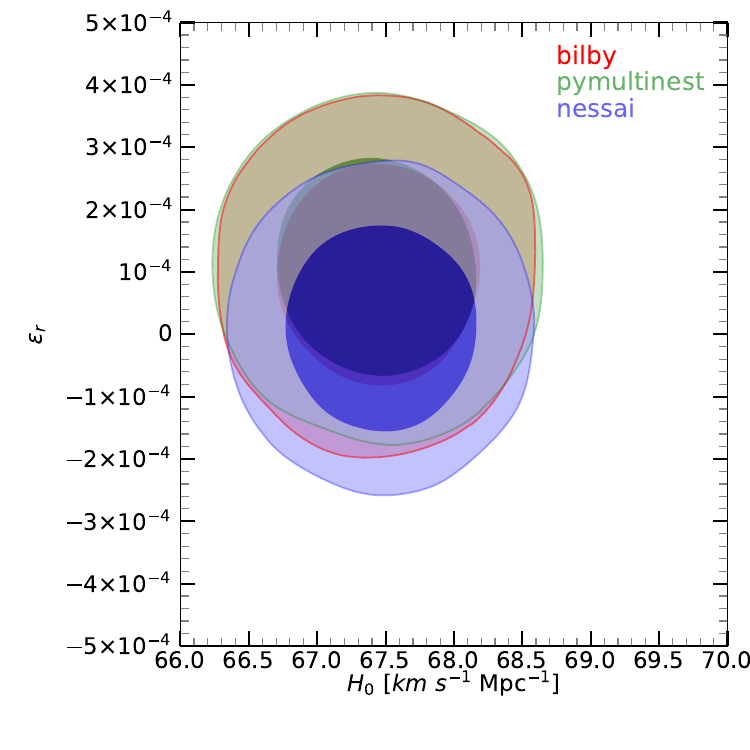}
\includegraphics[scale=0.5]{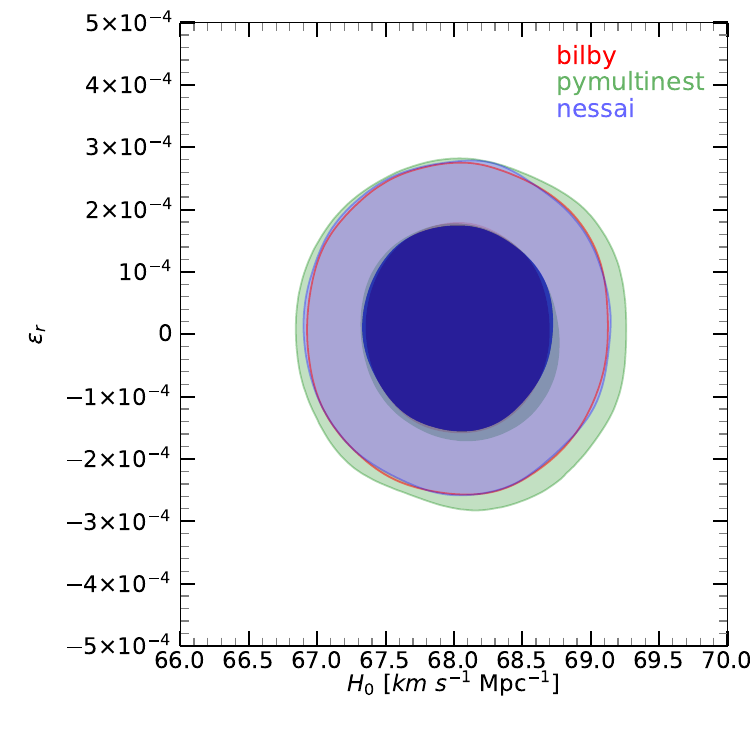}
\includegraphics[scale=0.5]{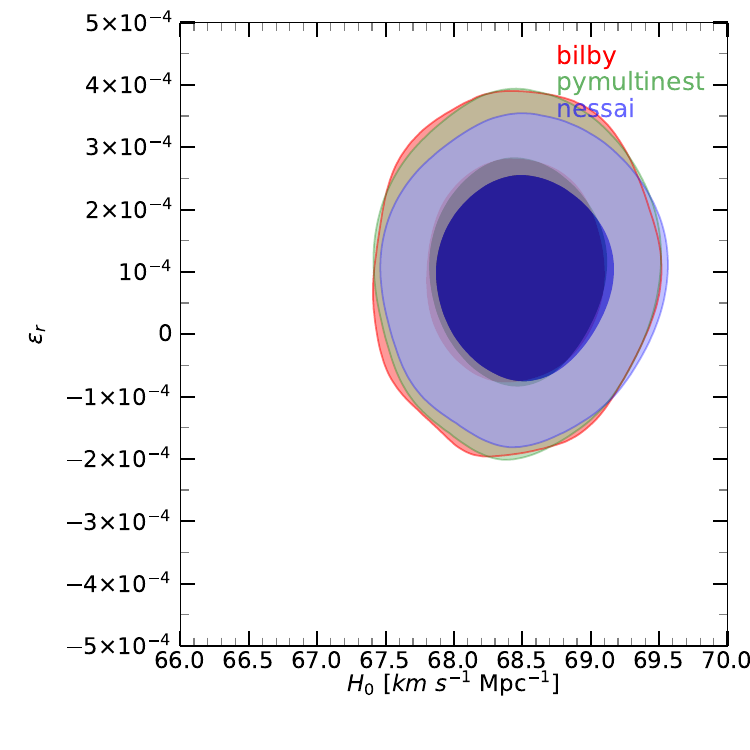}
\end{minipage}
	
\begin{minipage}[t]{1\paperwidth}%
\hspace{-1.2cm}\includegraphics[scale=0.5]{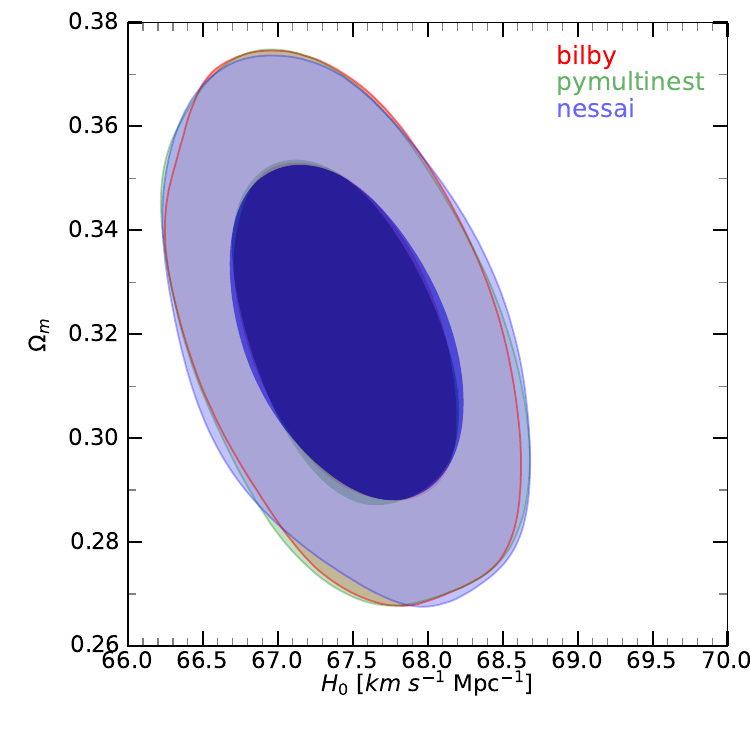}	
\includegraphics[scale=0.5]{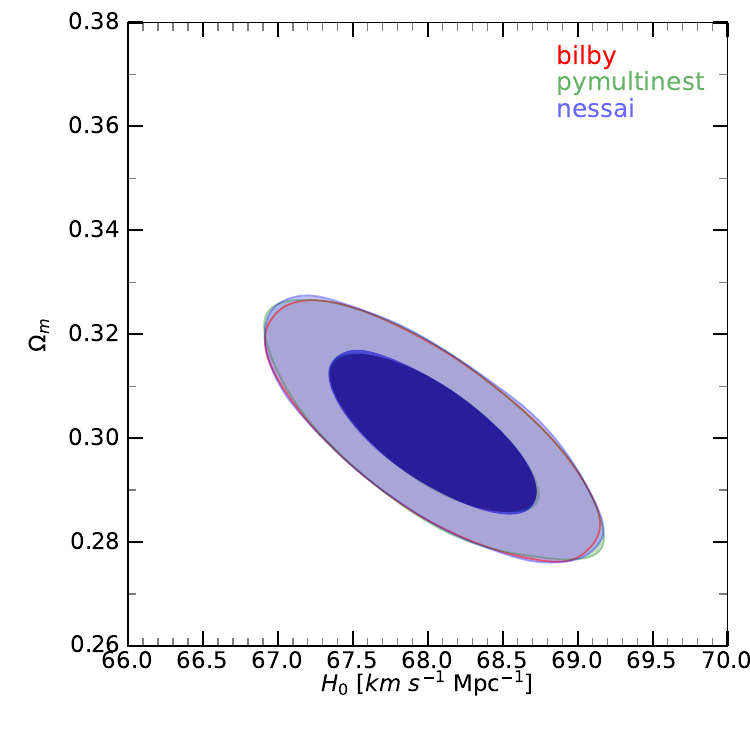}	
\includegraphics[scale=0.5]{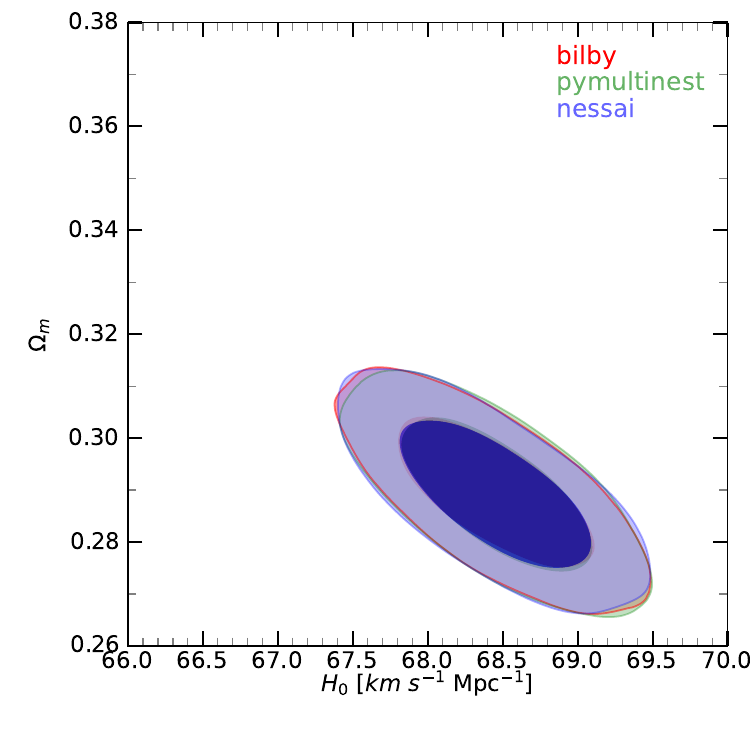}  
\end{minipage}  
\caption{\justifying 
Marginalized Bayesian posterior distributions, $H_{0}-\Omega_{m}$(bottom), $H_{0}-\epsilon_{r}$ (center), $H_{0}-n$ (top), corresponding to the grayscales on Fig.\ref{fig:result_fit_dist_CC}, of the model parameters for three sampler methods used in this paper, namely \texttt{bilby}, \texttt{nessai}, and \texttt{PyMultiNest}. From left to right using the CC observations measured with DA method; BAO dataset; combined sample CC and BAO datasets.
}
\label{fig:h2d_Bays_posterior_dist_H0_vs_omegaM}		
\end{figure*}

\begin{figure}	
	\includegraphics[scale=0.42]{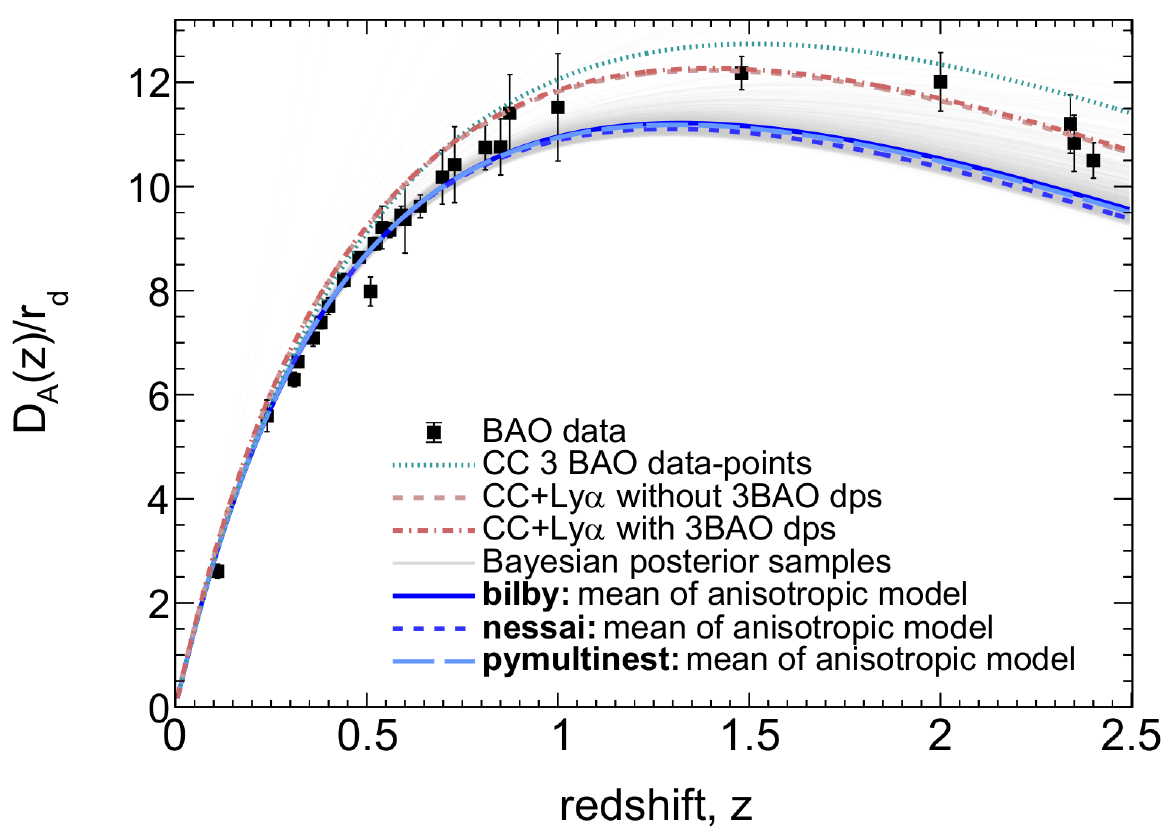}
	\includegraphics[scale=0.42]{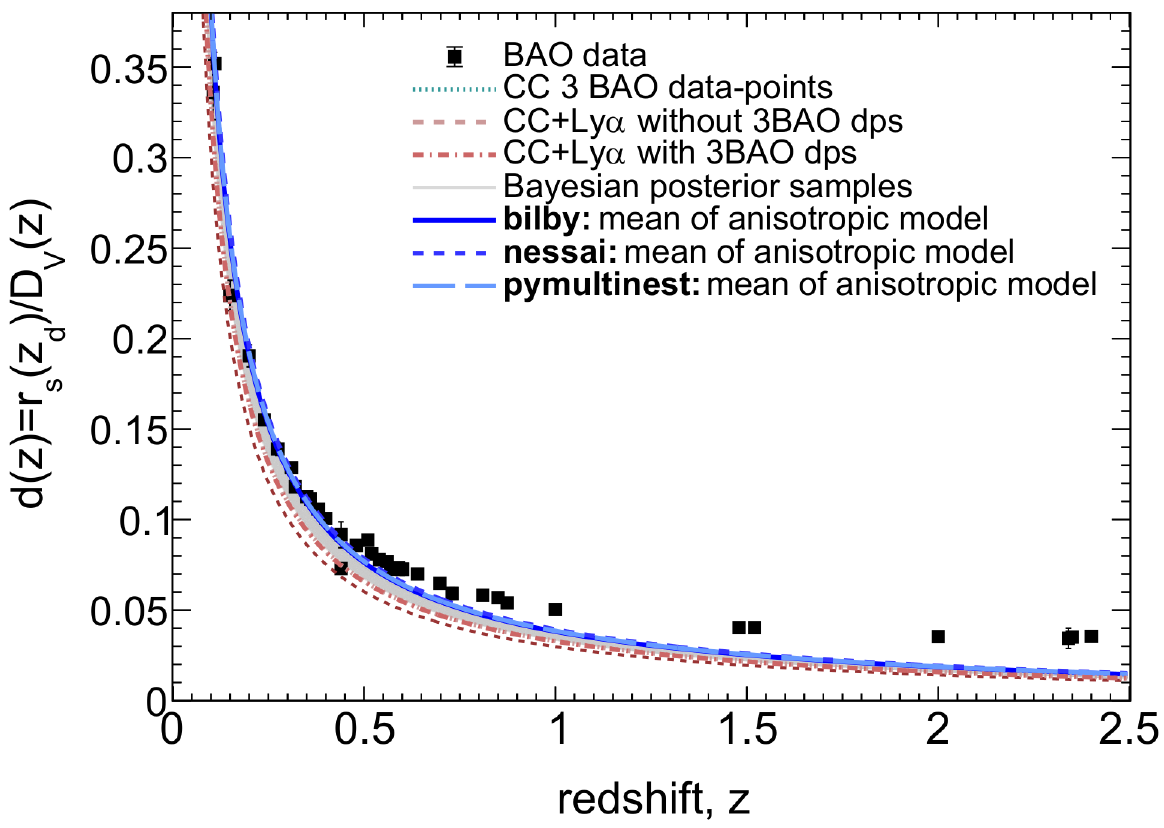}
	\caption{\justifying 
		The anisotropic BAO parameter, $D_{A}(z)/r_{d}$, vs. $z$ (top) and $r_{s}(z_{d})/D_{V}(z)$ vs. $z$ (bottom). The data points are taken from the compilation of BAO measurements from diverse releases, see Table~\ref{tab:zi_dzi_onlyBAO}. 
		The 
		red dotted line shows the $D_{A}(z)/r_{d}$ as function of redshift from the result parameters obtained from the Bayesian sampling to 3 galaxy distributions, see Fig.\ref{fig:result_fit_dist_CC}(a).
		The red dashed line shows the $D_{A}(z)/r_{d}$ as function of redshift from the result parameters obtained from the Bayesian sampling 
        to $H_{0}$ parameter observation measured with BAO method,         
        see Fig.\ref{fig:result_fit_dist_CC}(b); The red dot-dashed line shows the $D_{A}(z)/r_{d}$ as function of redshift from the result parameters obtained from the Bayesian sampling to CC data measured with DA method, see Fig.\ref{fig:result_fit_dist_CC}(c); 
		The blue solid line shows the $D_{A}(z)/r_{d}$ as function of redshift from the result parameters obtained from the Bayesian sampling to BAO distance measurements data with the help of the \texttt{bilby} method;
		The blue short-dashed shows the same, using  
		the \texttt{nessai} method;
		The blue long-dashed shows the same, using  
		the \texttt{PyMultiNest} method.
		The thin gray lines illustrates the Bayesian posterior sampling. 
	}
	\label{fig:result_fit_dist_BAO}		
\end{figure}

\begin{figure}
    \begin{minipage}[t]{1\textwidth}%
\hspace{-1cm}\includegraphics[scale=0.39]{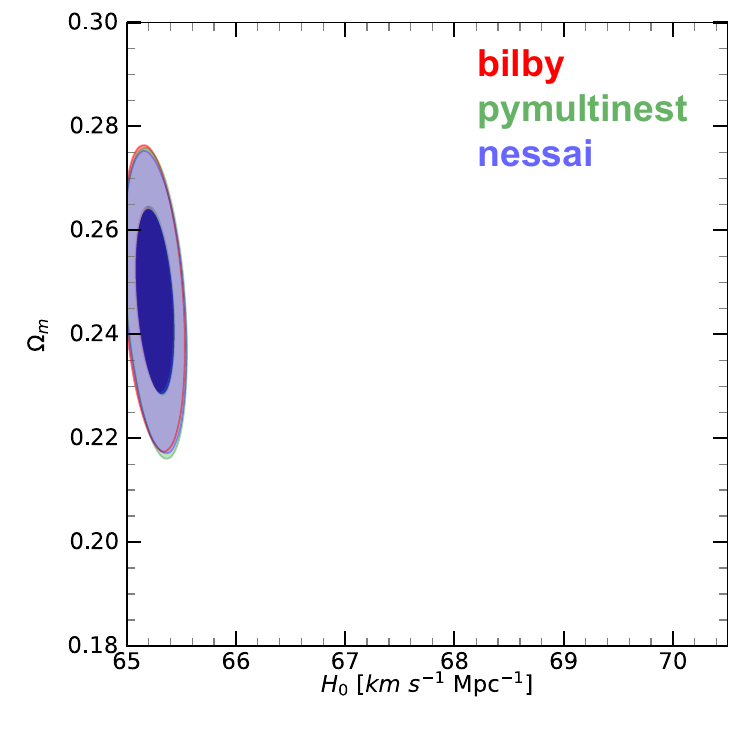}
	\includegraphics[scale=0.39]{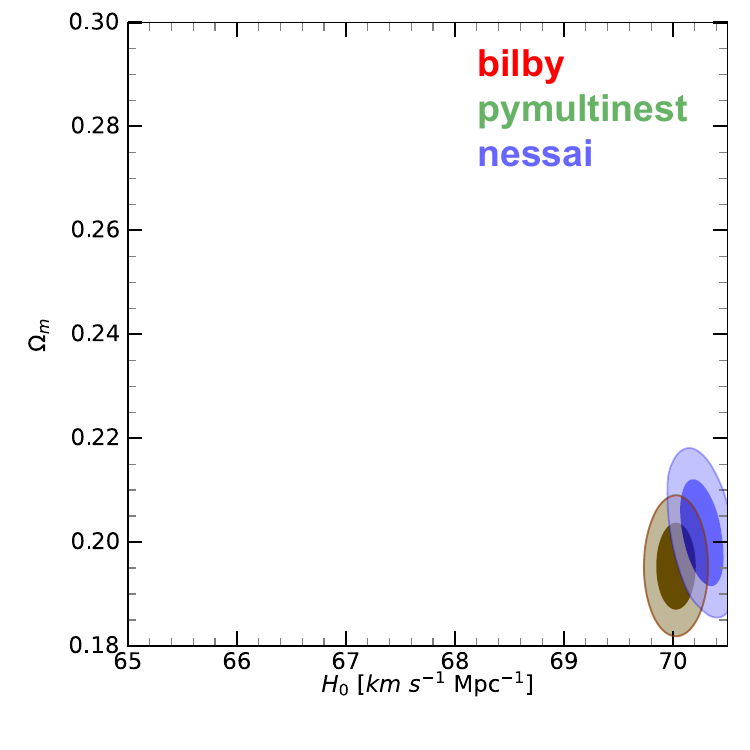}
\end{minipage}
	
	\begin{minipage}[t]{1\textwidth}%
\hspace{-1cm}\includegraphics[scale=0.39]{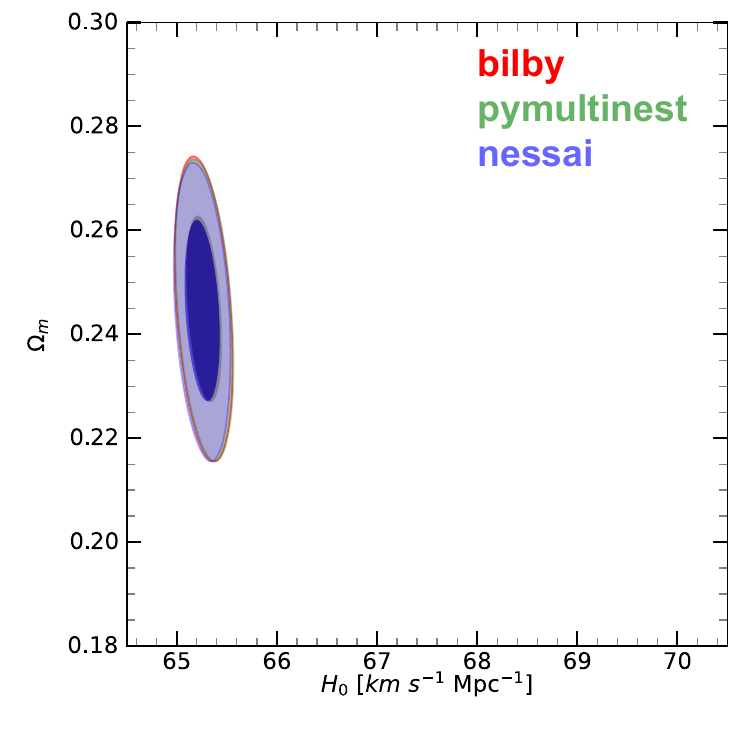}
	\includegraphics[scale=0.39]{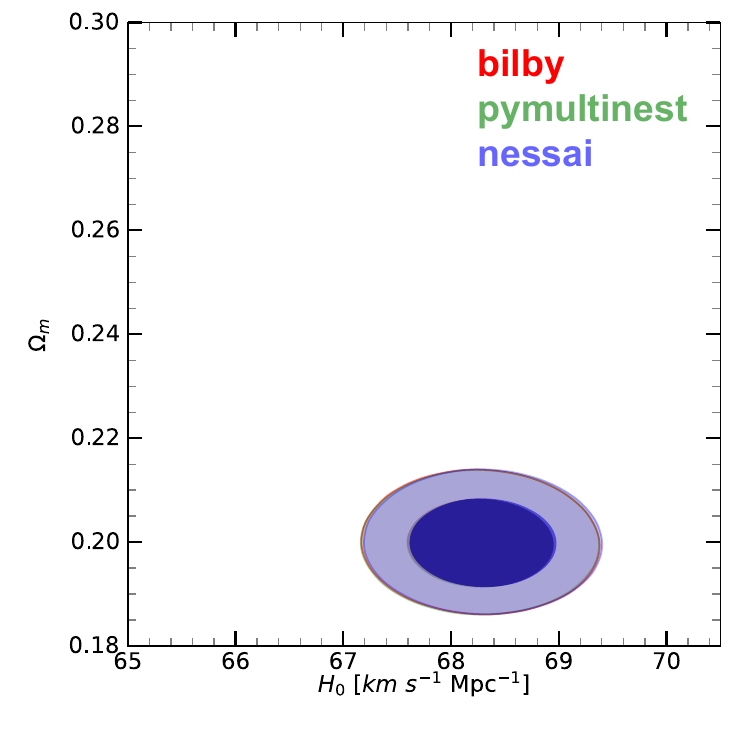}
\end{minipage}    
    \caption{\justifying 
		Two-dimensional marginalized posterior distributions for the anisotropic Hubble constant ($H_0^{\rm{anis}}$) and the matter density parameter ($\Omega_m^{\rm{anis}}$), derived from Bayesian analysis under different data, as shown on Fig.~\ref{fig:result_fit_dist_BAO}. 
		The four panels correspond to: 
		(top-left) BAO data in the $z - D_{A}(z)/r_{d}$ plane with uniform priors; 
		(top-right) the same BAO data with Gaussian priors; 
		(bottom-left) BAO data in the $z - r_s(z_d)/D_V(z)$ plane with uniform priors; 
		(bottom-right) BAO+DA dataset with Gaussian priors. 
		Each plot shows combined results from three samplers (\texttt{bilby}, \texttt{PyMultiNest}, and \texttt{nessai}), illustrating the influence of prior selection and data combination on the joint constraints of the cosmological parameters. 
		The contours represent 68\% and 95\% confidence  
		intervals.
	}
	\label{fig:result_BAOdzi-Dard_Bays_posterior_dist_H0_vs_omegaM}		
\end{figure}

\subsubsection{Baryon Acoustic Oscillation data}
\label{sec:BAO_data}

Baryonic Acoustic Oscillations (BAO) serve as a vital tool in cosmology, enabling us to probe the 
structure of the Universe on a large 
scale. These fluctuations
originate from acoustic waves that propagated through the early Universe, causing the compression of baryonic matter and radiation within the photon-baryon fluid. 
Therefore BAO measurements are useful to study the angular-diameter distance as a function of redshift 
and the evolution of the Hubble parameter. These measurements are represented by using angular scale and redshift separation. 

On large angular scales, baryon acoustic oscillations occur as separate peaks and are thought to be pressure waves caused by cosmic perturbations in the baryon-photon
plasma during the recombination era (BOSS) \citep{Blake:2011en, SDSS:2009ocz}.

They are commonly written in terms of the dimensionless ratio
\begin{equation}
d(z)=\frac{r_{s}(z_{d})}{D_{V}(z)},
\label{eq:BAO_dz}
\end{equation}
where the spherically-averaged distance reads
\begin{equation}
D_{V}(z)\equiv [zD_{M}^{2}(z) D_{H}(z)]^{1/3}.
\label{eq:BAO_DV}
\end{equation}
The powers 
$2/3$ and $1/3$ approximately account for two transverse and one radial dimension and the extra factor 
$z$ stems from 
conventional normalization. $D_{H}(z)$ is the Hubble distance at redshift $z$; $D_{M}(z)$ is the (comoving) angular diameter distance, which depends on the expansion history and curvature.

$r_{s}(z_{d})$ represents the comoving size of the sound horizon at the drag redshift, $z_{d}= 1059.6$ \citep{Planck:2015fie}:
\begin{equation}
r_{s}(z_{d})=\int_{z_{d}}^{\infty} \frac{c_{s}}{H(z)}dz.
\label{eq:BAO_rs_zd}
\end{equation}
Here, $c_{s}=1/\sqrt{3(1+\mathcal{R})}$ represents the sound speed of the baryon-photon fluid, and $\mathcal{R}=\frac{3\Omega_{b_{0}}}{4\Omega_{r_{0}(1+z)}}$ with
$\Omega_{b_{0}}=0.022 h^{-2}$ \citep{Cooke:2016rky} and $\Omega_{r_{0}}=\Omega_{\gamma_{0}}\left( 1 + \frac{1}{8} \left(\frac{4}{11}\right)^{\frac{4}{3}} N_{\rm{eff}} \right) $ and where $\Omega_{\gamma_{0}}=2.469\times 10^{-5} h^{-2}$ and $N_{\rm{eff}}=3.046$ \citep{Dodelson:2003ft}.

The expressions utilized for non-correlated BAO data are as follows,
\begin{equation}
	\chi^{2}_{BAO/\rm{noncov}} = \sum_{i=1}^{30} \frac{[H_{\rm{obs}}(z_{i,\theta})- H_{\rm{th}}^{BAO}(z_{i})]^{2} }{\sigma^{2}_{H_{\rm{obs}}(z_{i})}},
	\label{eq:chi2_BAO}
\end{equation}
In these expressions, $H_{\rm{th}}^{BAO}(z_{i})$ denotes the theoretical values
of the Hubble parameter for a particular model characterized by model parameters $\theta$. Conversely, $H_{\rm{obs}}^{BAO}(z_{i})$ corresponds to the observed Hubble parameter acquired through the BAO method, while $\sigma_{H_{\rm{obs}}(z_{i})}$ represents the experimental uncertainty associated with the observed values of $H_{\rm{obs}}^{BAO}$.

In order to extract the most information from BAO observations, it is common to analyze them in different parametrizations that emphasize different aspects of the distance-redshift relation. In particular, we consider fits to both the $z - r_s(z_d)/D_V(z)$ and the $z - D_A(z)/r_d$ planes. The former, based on the volume-averaged distance $D_V(z)$, offers a spherically averaged constraint that is robust to uncertainties in radial versus angular contributions. The latter, based on the angular diameter distance $D_A(z)$, provides direct access to the transverse clustering scale and is especially useful when combined with anisotropic (2D) measurements. By analyzing both planes, we gain a more comprehensive and robust constraint on the underlying cosmological parameters, especially the expansion history encoded in $H(z)$ and the geometry-sensitive quantity $D_A(z)$. Such decomposition has been widely used in large-scale structure analyses, including those from BOSS \citep{BOSS:2016wmc}, eBOSS \citep{eBOSS:2020yzd}, and DESI forecasts \citep{DESI:2016fyo}.

This approach allows us to explore potential tensions or consistencies between different types of BAO measurements, as well as test the sensitivity of our model to radial versus angular distance scales. It also enhances our ability to constrain anisotropic cosmologies or deviations from standard $\Lambda$CDM.

We consider a compilation of 35 $D_{V}(z)$ measurements and 29 the angular diameter distance $D_{A}(z)$ measurements as shown in Table \ref{tab:zi_dzi_onlyBAO}. Therefore, chi-squared functions can be computed for the $D_{V}(z)$ and  $D_{A}(z)$ measurements, denoted by 
\begin{equation}
\chi^{2}_{D_{V}} = \sum_{i=1}^{35} \frac{[(r_{d}/D_{V})^{\rm{obs}}(z_{i})- (r_{d}/D_{V})^{\rm{th}}(z_{i})]^{2} }{\sigma^{2}_{d_{i}}(z_{i})},
\label{eq:chi2_BAO_DV}
\end{equation}
and
\begin{equation}
	\chi^{2}_{D_{A}} = \sum_{i=1}^{29} \frac{[(D_{A}/r_{d})^{\rm{obs}}(z_{i})- (D_{A}//r_{d})^{\rm{th}}(z_{i})]^{2} }{\sigma^{2}_{D_{A}/r_{d}}(z_{i})},
	\label{eq:chi2_BAO_DA}
\end{equation}
respectively.

Figure \ref{fig:result_fit_dist_CC} shows the fits to the DA and BAO data samples, the thin gray lines illustrate the Bayesian posterior sampling, solid black lines show the best-fit, while Figure \ref{fig:h2d_Bays_posterior_dist_H0_vs_omegaM} displays the recovered cosmological parameters for the model under examination. 
The Hubble constant, $H_{0}$, values range from 67.44 to 68.54 $\rm \,km\,s^{-1}\,Mpc^{-1}$ across samplers and data subsets. The combined DA and BAO dataset gives tighter constraints, with $H_{0} \approx 68.53 - 68.54 \rm \,km\,s^{-1}\,Mpc^{-1}$, uncertainties of 0.4-0.5, close to the standard cosmological value \citep[e.g., Plank 2018,][]{Planck:2018vyg}. 
The same is true for the matter density parameter $\Omega_{m}$, as a combined DA and BAO dataset suggests $\Omega_{m} \approx 0.29 - 0.33$, which is consistent with the standard cosmology. $\epsilon_{r}$ is consistently small $\approx 10^{-5} - 10^{-4}$, indicating low anisotropy. The \texttt{nessai} sampler often estimates a lower $\epsilon_{r}$. $n$ is poorly constrained by the BAO and DA datasets. 
All results of cosmological anisotropic model parameters are shown in Table~\ref{tab:results}.

Figure \ref{fig:result_fit_dist_BAO} shows the fits to the anisotropic BAO parameter, $D_{A}(z)/r_{d}$, vs. $z$ (top) and $r_{s}(z_{d})/D_{V}(z)$ vs. $z$ (bottom). The data points are taken from the compilation of BAO measurements from various releases; see Table~\ref{tab:zi_dzi_onlyBAO}. 
The red dotted line shows the $D_{A}(z)/r_{d}$ as a function of redshift from the result parameters obtained from the Bayesian sampling to 3 galaxy distributions; see Fig.\ref{fig:result_fit_dist_CC}(a).
The red dashed line shows the $D_{A}(z)/r_{d}$ as function of redshift from the result parameters obtained from the Bayesian sampling to BAO data, see Fig.\ref{fig:result_fit_dist_CC}(b); The red dot-dashed line shows the $D_{A}(z)/r_{d}$ as function of redshift from the result parameters obtained from the Bayesian sampling to CC data measured with DA method, see Fig.\ref{fig:result_fit_dist_CC}(c); 
The blue solid line shows the $D_{A}(z)/r_{d}$ as function of redshift from the result parameters obtained from the Bayesian sampling to BAO distance measurements data with the help of the \texttt{bilby} method. 
The blue short-dashed shows the same, using the \texttt{nessai} method;
The blue long-dashed shows the same, using the \texttt{PyMultiNest} method.
The thin gray lines illustrate the Bayesian posterior sampling. 
Bold solid, dashed and short dashed lines show the best-fit. 
    
Figure \ref{fig:result_BAOdzi-Dard_Bays_posterior_dist_H0_vs_omegaM} displays the recovered cosmological parameters for the anisotropic model under examination. 
In the $z-r_s(z_{d})/D_{V}(z)$ plane, the Hubble constant is estimated as $H_0^{\text{anis}} \approx 71.00 - 75.52 \rm \,km\,s^{-1}\,Mpc^{-1}$ with tight uncertainties ($\pm 0.001 - 0.48$) using uniform ($\mathcal{U}$) priors, suggesting higher $H_0$ values compared to the Cosmic Chronometers (CC) parameter plane. However, these cases yield extremely high $\chi^2/N_{\text{dof}} \approx 348.31 - 664.08$, indicating poor model compatibility. With Gaussian ($\mathcal{G}$) priors, $H_0^{\text{anis}} \approx 71.00-71.26 \rm \,km\,s^{-1}\,Mpc^{-1}$, with slightly better fits ($\chi^2/N_{\text{dof}} \approx 348.31$). By assuming a wider uncertainties ($10\sigma_{\text{exp}}$), we get $H_0^{\text{anis}} \approx 69.17-72.22 \rm \,km\,s^{-1}\,Mpc^{-1}$, with significantly improved fits ($\chi^2/N_{\text{dof}} \approx 2.87-3.44$). In the $z-D_A(z)/r_d$ plane, $H_0^{\text{anis}} \approx 65.10-65.11 \rm \,km\,s^{-1}\,Mpc^{-1}$ for $\mathcal{U}$ priors and $H_0^{\text{anis}} \approx 65.82-66.12 \rm \,km\,s^{-1}\,Mpc^{-1}$ for $\mathcal{G}$ priors, both aligning more closely with CC values and showing good fits ($\chi^2/N_{\text{dof}} \approx 1.79-2.97$). Shifted priors (see discussion in Sec.~\ref{sec:discussion}) yield $  H_0^{\text{anis}} \approx 65.10-72.16 \rm \,km\,s^{-1}\,Mpc^{-1}$, but fits deteriorate ($  \chi^2/N_{\text{dof}} \approx 1.79-28.62  $).

The matter density parameter, $\Omega_m^{\text{anis}}$, shows variability in the $z-r_s(z_{d})/D_{V}(z)$ plane, with low values ($\sim 0.20-0.23$) and tight uncertainties, indicating potential model tension, especially in poor-fit cases. In the $z-D_A(z)/r_d$ plane, $\Omega_m^{\text{anis}} \approx 0.21-0.33$, aligning better with CC values ($\sim 0.3$), particularly with $\mathcal{U} $ priors ($\chi^2/N_{\text{dof}} \approx 1.79$). The model struggles to consistently constrain $\Omega_m^{\text{anis}}$ across BAO parameter planes, with shifted priors compounding variability.

The recombination anisotropy 
amplitude, $\epsilon_r$, varies significantly. In the $z-r_s(z_{d})/D_{V}(z)$ plane with standard $\mathcal{U}$ and $\mathcal{G}$ priors, $\epsilon_r \approx 1.88-4.84 \times 10^{-4}$, with large uncertainties ($ \sim \pm 10^{-4}$) and poor fits ($\chi^2/N_{\text{dof}} \approx 348-664$), suggesting forced anisotropy parameters compensate for data mismatches, or model mismatch.  
With $10\sigma_{\text{exp}}$ $\mathcal{U}$ priors, $\epsilon_r \approx 5.52-5.82 \times 10^{-6}$, with tight uncertainties and good fits ($\chi^2/N_{\text{dof}} \approx 3.37-3.44$), indicating minimal anisotropy. In the $z-D_A(z)/r_d$ plane, $\epsilon_r \approx 9.94-10.2 \times 10^{-6}$ for $\mathcal{U}$ priors with very tight uncertainties ($\sim \pm 10^{-7}$) and good fits ($\chi^2/N_{\text{dof}} \approx 1.79 - 1.80$), while $\mathcal{G}$ priors yield $\epsilon_r \approx 4.85-5.05 \times 10^{-4} $ with worse fits ($\chi^2/N_{\text{dof}} \approx 2.65-2.97$). Shifted priors produce $\epsilon_r \approx 9.34 \times 10^{-6}-6.03 \times 10^{-4}$, with fits ranging from good to poor ($\chi^2/N_{\text{dof}} \approx 1.79-28.62$). Thats suggests that restrictive priors may artificially inflate anisotropy. The model generally indicates detectable but small anisotropy, with better fits corresponding to lower $\epsilon_r$. \texttt{bilby} and \texttt{PyMultiNest} are highly consistent, providing robust estimates, while \texttt{nessai} shows deviations, sometimes worse fits, possibly due to exploring complex posterior regions. More detailed results are shown in Table~\ref{tab:results}.

\begin{figure}
	\includegraphics[scale=0.45]{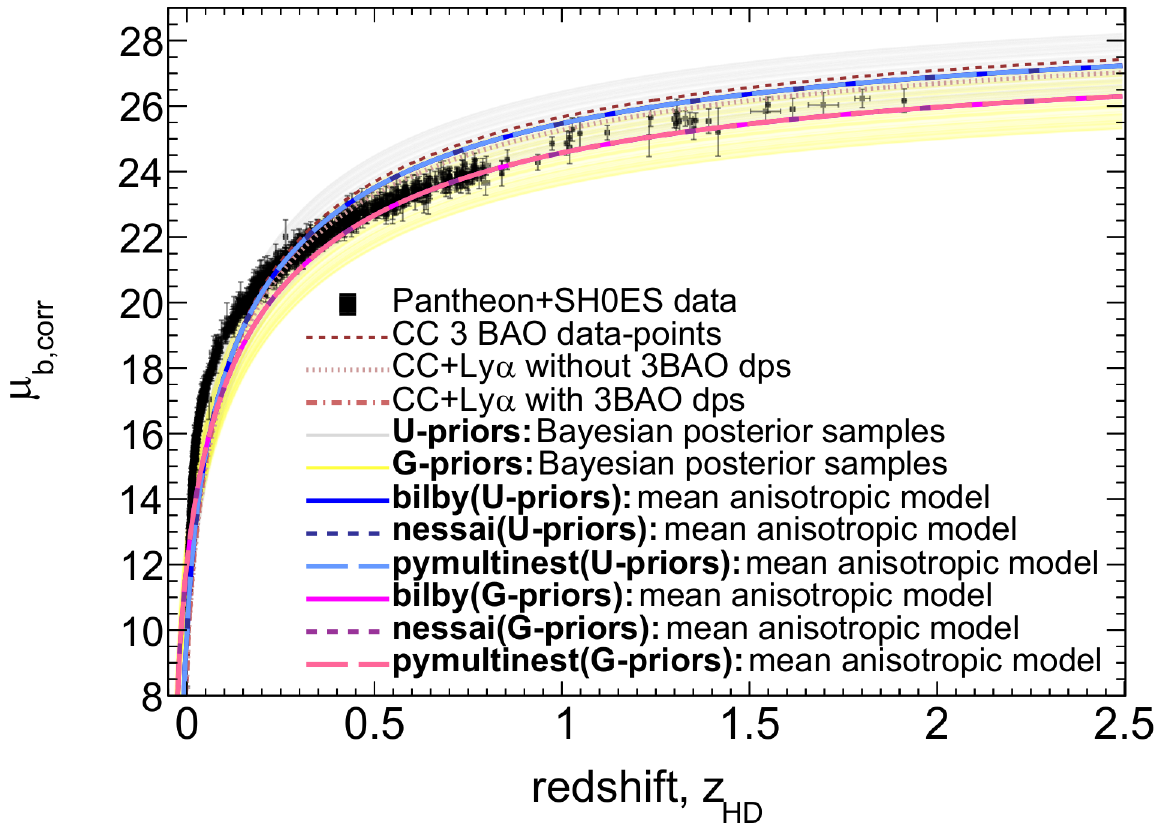}
	\caption{\justifying 
		The luminosity distance parameter, $\mu$, as function of redshift, $z$. The data points are taken from \citep{Scolnic:2021amr}, we refer to {\texttt{\text{m\_b\_corr}}} column in this dataset. 
		The blue solid line shows the $\mu$, see Eq.\eqref{eq:muOfT}, as function of redshift from the result parameters obtained from the Bayesian sampling to Pantheon+SHOE data of the distance modulus measurements data 
		with the help of the \texttt{bilby} method;
		The blue short-dashed displays the same, using
		the \texttt{nessai} method;
		The blue long-dashed shows the same, using
		the \texttt{PyMultiNest} method.
		The thin gray lines illustrates the Bayesian posterior sampling using $\mathcal{U}$ priors, while thin yellow lines - $\mathcal{G}$ priors.				
		Solid blue line show the best-fit within \texttt{bilby} sampler assuming $\mathcal{U}$ priors,
		solid red line show the best-fit within \texttt{bilby} sampler for $\mathcal{G}$ priors.
	}
	\label{fig:result_fit_dist_mu_dist}	
\end{figure}

\begin{figure}
\begin{minipage}[t]{1\textwidth}%
\hspace{-0.8cm}\includegraphics[scale=0.39]{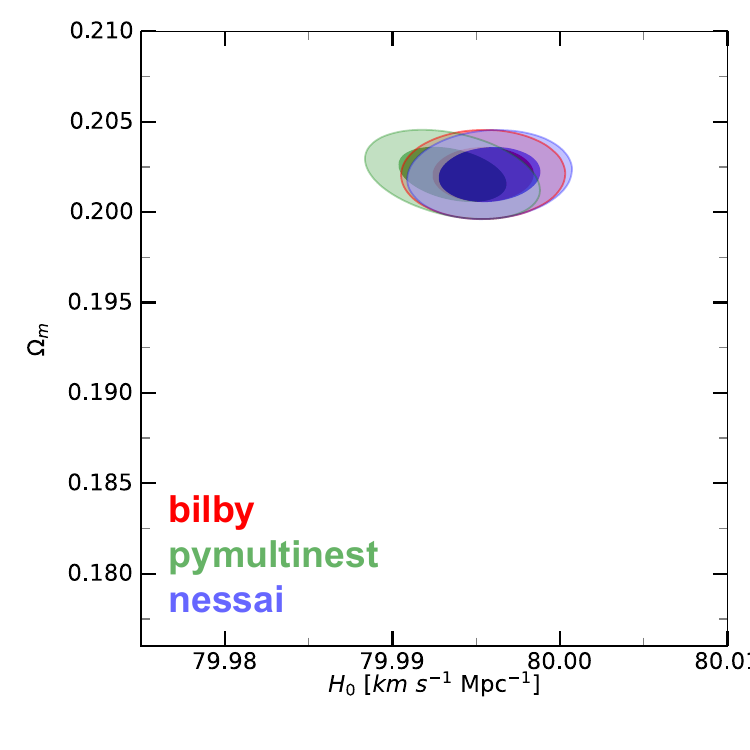}
\includegraphics[scale=0.39]{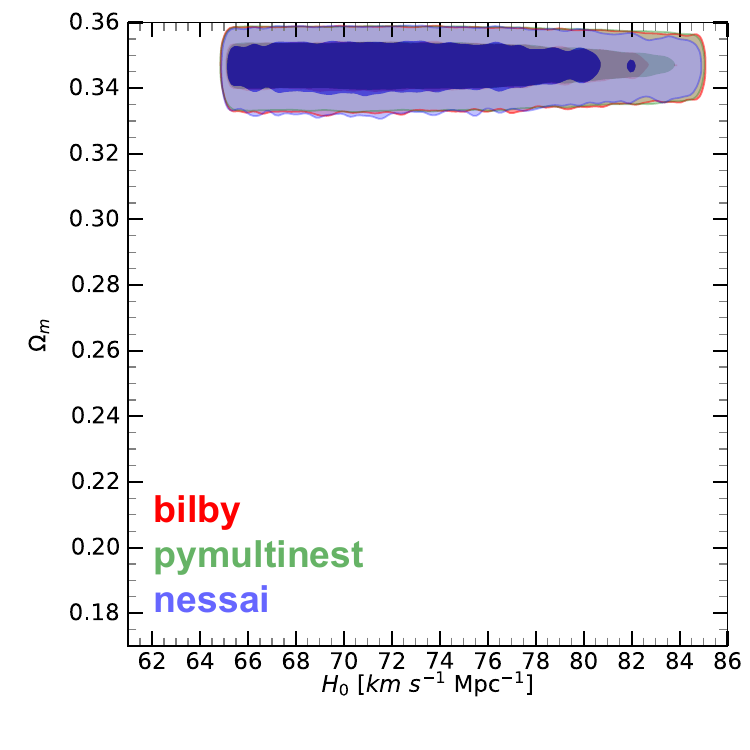}
\end{minipage}
       
\begin{minipage}[t]{1\textwidth}%
\hspace{-0.8cm}\includegraphics[scale=0.39]{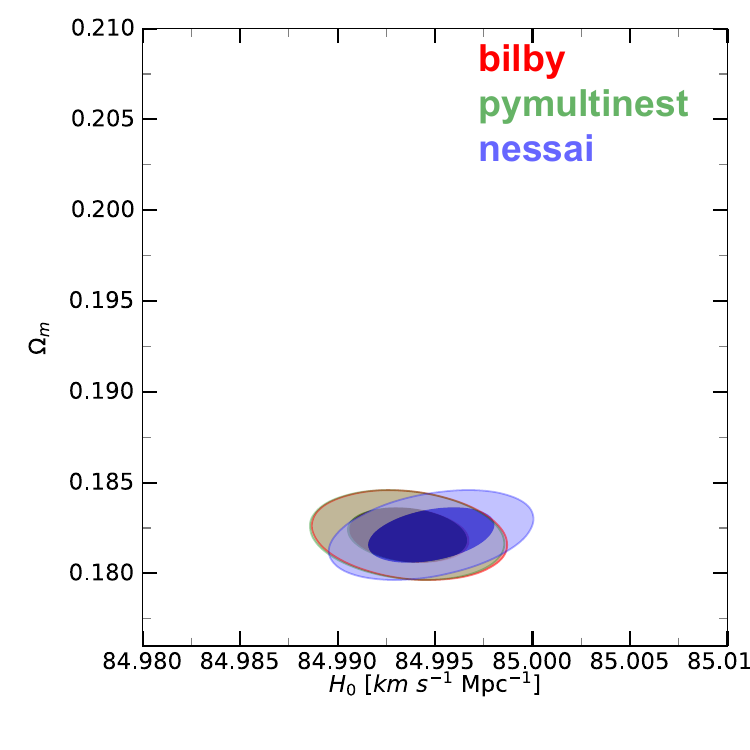}
\includegraphics[scale=0.39]{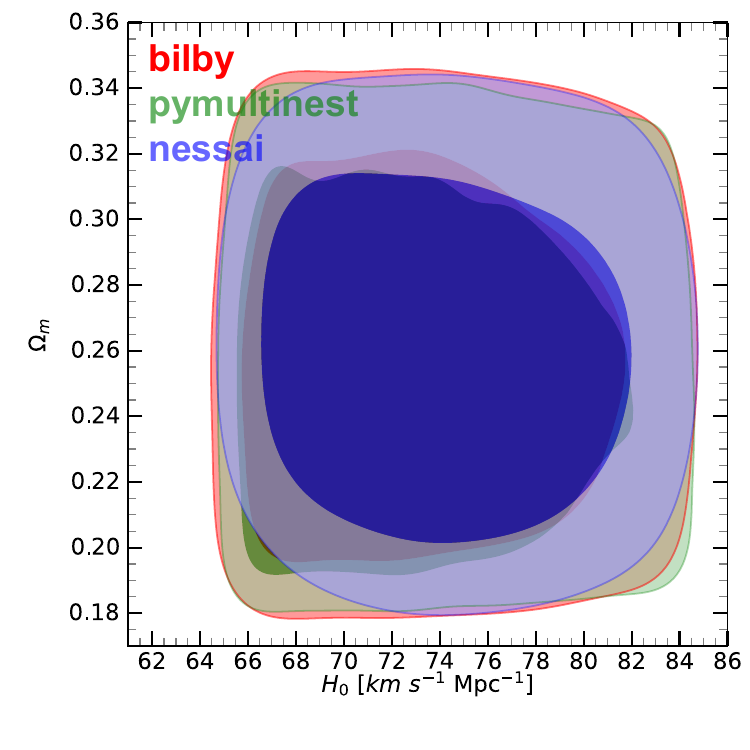} 
\end{minipage}
\caption{\justifying 
Two-dimensional marginalized posterior distributions for the Pantheon+SHOE dataset, as shown on Fig.~\ref{fig:result_fit_dist_mu_dist}, of the model parameters 
for three sampler methods used in this paper, namely \texttt{bilby}, \texttt{nessai}, and \texttt{PyMultiNest}. 
The panels correspond to: (left) fit results with uniform priors, where we do fix (top) or 
release (bottom) the $H_{0}$ parameter during the Bayesian sampling; (right) fit results with Gaussian priors. 
The contours represent 68\% and 95\% confidence intervals.
}
\label{fig:h2d_PantheonSHOE_Bays_posterior_dist_H0_vs_omegaM}		
\end{figure}

\subsubsection{Pantheon and Supernova Cosmology Project SNe type Ia data}
\label{sec:Pantheon_data}

The Pantheon sample is a combination of five subsamples: 
PS1, SDSS, SNLS, low-$z$, and HST that gives the largest supernovae sample of 1048 measurements, spanning over the redshift range: $0.01 < z < 2.3$ \citep{Pan-STARRS1:2017jku}. 

For the SNIa data, we have used the Pantheon supernovae sample \citep{Pan-STARRS1:2017jku}, which consists of five subsamples: PS1, SDSS, SNLS, low-$z$, and HST that gives the largest supernovae sample of 1048  
spectroscopically confirmed SNe Ia covering the redshift range: $0.01 < z < 2.3$ \citep{Pan-STARRS1:2017jku}. 
The distribution of SNe Ia in Pantheon are inhomogeneous, and half of them are located in the south-east of the galactic coordinate system. The systematic uncertainties were reduced by cross-calibration between subsamples in Pantheon. Therefore, the Pantheon sample could bring a much stronger constraint for the anisotropy of the Universe. 
Compared to the 753 supernovae  
of SCP Union2.1, or the 53 supernovae  
of \citep{SupernovaCosmologyProject:2011ycw}, 
the number of SNe Ia in the Pantheon sample is enlarged. 

Over recent years, several compilations of Type Ia supernova data, such as SCP Union2 \citep{Amanullah:2010vv}, SCP Union2.1 \citep{SupernovaCosmologyProject:2011ycw}, JLA \citep{SDSS:2014iwm}, Pantheon \citep{Pan-STARRS1:2017jku}, and the latest addition, Pantheon+ \citep{Scolnic:2021amr} have been gathered.

	Figure \ref{fig:result_fit_dist_mu_dist} shows the fits to the luminosity distance parameter, $\mu$, as function of redshift, $z$. The data points are taken from Ref.~\citep{Scolnic:2021amr}
     and, more precisely, we refer to the {\texttt{\text{m\_b\_corr}}} column in this dataset. 	
	The solid lines shows the $\mu$, from  
    Eq.\eqref{eq:muOfT}, as function of redshift from the result parameters obtained from the Bayesian sampling to Pantheon+SHOE data \citep{Scolnic:2021amr} of the distance modulus measurements data with the help of the \texttt{bilby} method;	
	The blue short-dashed displays the same, using the \texttt{nessai} method;
	The blue long-dashed shows the same, using the \texttt{PyMultiNest} method.
	The thin gray lines illustrates the Bayesian posterior sampling using the $\mathcal{U}$ priors, while the thin yellow lines - the $\mathcal{G}$ priors. 
	The solid  
    blue line show the best-fit within the \texttt{bilby} sampler assuming the $\mathcal{U}$ priors, 
	the solid red line show the best-fit within the \texttt{bilby} sampler for the $\mathcal{G}$ priors.
	
	Figure \ref{fig:h2d_PantheonSHOE_Bays_posterior_dist_H0_vs_omegaM} displays the recovered cosmological parameters. 
	Assuming the $\mathcal{U}$ priors we see that, for the SHOE dataset, the model suggests the Hubble constant, $H_{0} \approx 65.00 - 65.03 \rm \,km\,s^{-1}\,Mpc^{-1}$ with very tight uncertainties ($\pm 0.01-0.02$). These values appear consistent across Bayesian samples, but remain lower that the SHOE reported value  ($73 \rm \,km\,s^{-1}\,Mpc^{-1}$). For the Pantheon+SHOE data, we get $H_{0}\approx 80.0$ with tiny uncertainty - this is very high compared to the Plank or SHOE reported values. This behavior may reflect the fact that the Pantheon's supernovae data is responsible for the increase in the $H_0$ fitted values (see also Table~\ref{tab:results_muD})
    . If one will assume $\mathcal{G}$ priors, the model suggests $H_{0} \approx 72.80 - 74.07 \rm \,km\,s^{-1}\,Mpc^{-1}$ which is closer to the SHOES measurements with large uncertainties ($\pm 5.02-6.56$). Similar values were obtained for the Pantheon+SHOE data, that is 
	$H_{0} \approx 72.85 - 73.55 \rm \,km\,s^{-1}\,Mpc^{-1}$ with uncertainties ($\pm 5.00-6.55$). With 
    the above
    , one may assume that the $\mathcal{G}$ priors help to stabilize the $H_{0}$ values.
			
	The matter density parameter, $\Omega_{m}$, shows consistency  
    with standard cosmology values, $\Omega_{m}\approx 0.35$, but the $\mathcal{U}$ priors yield poor fits. Pantheon+SHOE with the $\mathcal{U}$ priors gives low $\Omega_{m}\approx 0.18-0.20$. The $\mathcal{G}$ priors stabilize $\Omega_{m}\approx 0.25-0.35$ with better fits with respect to the $\mathcal{U}$ priors choice.
	
	$\epsilon_{r}$ varies significantly, with the $\mathcal{U}$ priors giving higher values
	($\epsilon_{r} \approx 0.04-0.09$), suggesting stronger anisotropy but poor fits.
	The $\mathcal{G}$ priors yield lower $\epsilon_{r} \approx 0.002-0.03$, indicating minimal anisotropy and better fits, especially for the Pantheon+SHOE case. 
	$n$ is poorly constrained in all cases, ranging from $\sim$~1.77 to 3.12.
    All results are shown in Table~\ref{tab:results_muD}.

\begin{figure*}	
\begin{center}	

\begin{minipage}[t]{1\textwidth}%
\includegraphics[scale=0.295]{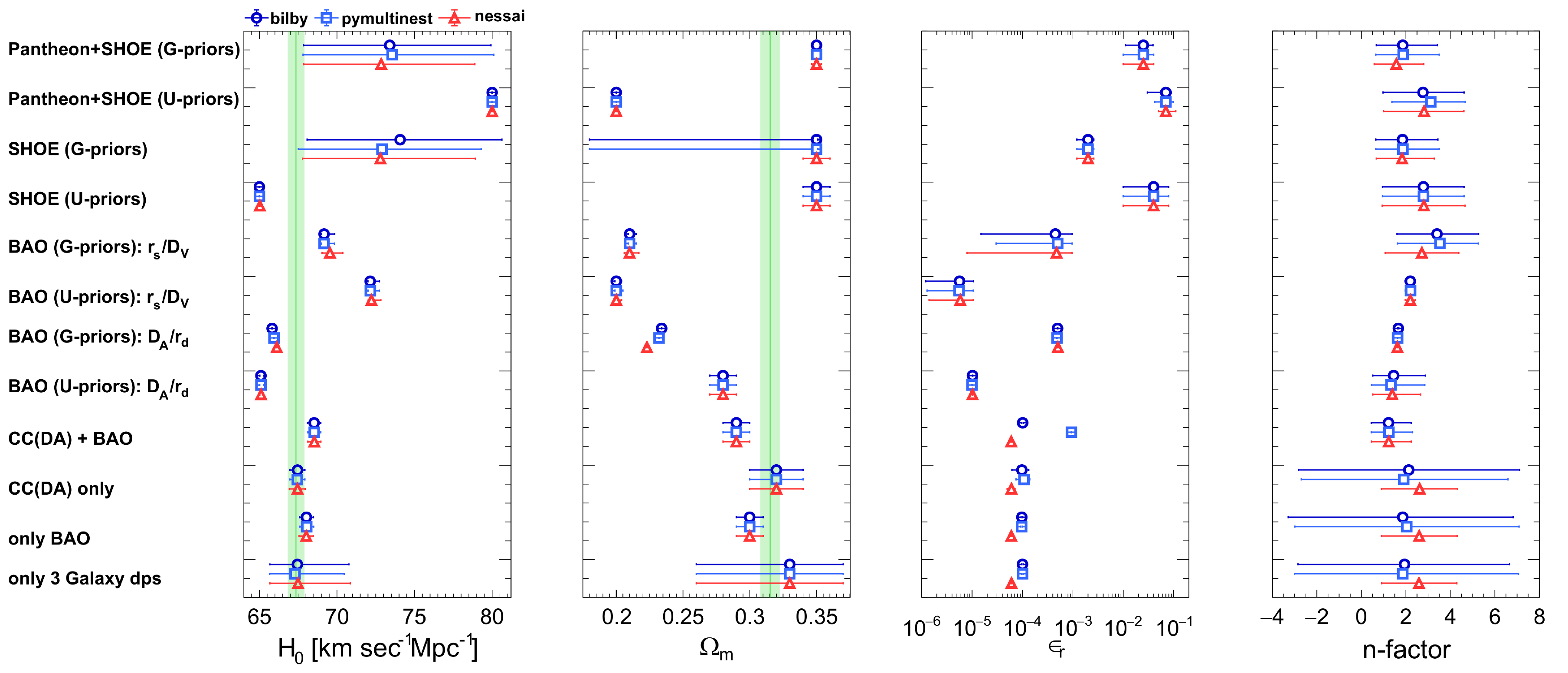}
\end{minipage}
\caption{\justifying 
	Comparison of the model parameters 
    estimates from various datasets and Bayesian inference samplers. Each row corresponds to a different dataset, including constraints from cosmic chronometers (CC), baryon acoustic oscillations (BAO), the SHOE dataset, and the Pantheon+SHOE dataset, analyzed with either Gaussian priors ($\mathcal{G}$ priors) or uniform priors ($\mathcal{U}$ priors). The three markers—circles, squares, and triangles—represent posterior mean values of $H_{0}$(left) and $\Omega_{m}$(right) obtained using the \texttt{bilby}, \texttt{PyMultiNest}, and \texttt{nessai} samplers, respectively. Horizontal bars denote the corresponding 68\% confidence 
	intervals. The green band indicates the reference value for $H_{0}$ and $\Omega_{m}$, from Planck 2018 \citep{Planck:2018vyg}. This figure demonstrates the sampler robustness and dataset dependence of inferred $H_{0}$ and $\Omega_{m}$ values across different modeling assumptions.
	}
 \label{fig:results_comb}	
\end{center}
\end{figure*}

\begin{table*}
	\caption{\justifying 
Posterior estimates of cosmological anisotropic model parameters using different 
Bayesian samplers and datasets. The parameters include the Hubble constant $H_{0}^{\rm{anis}}$, matter density parameter $\Omega_{m}^{\rm{anis}}$, energy transfer amplitude $\epsilon_{r}$, model parameter $n$, and the reduced chi-square statistic $\chi^{2}/N_{\rm{dof}}$. Results are shown for Hubble parameter observations measured with the help of BAO and DA methods.
For BAO dataset we show results for two different parameters $z-r_{s}(z_{d})/D_{V}(z)$ 
and $z-D_{A}(z)/r_{d}$ planes. Parameter estimates include 68\% confidence  
intervals. Notable differences are observed between samplers, especially for BAO dataset within $z-r_{s}(z_{d})/D_{V}(z)$ plane, where some fits return extremely large $\chi^{2}$ values, indicating poor model compatibility.
	}
	\centering
	\scalebox{0.95}{		
		\begin{tabular}{l|l|c|c|c|c|c}			        
			\hline    
			\hline            
			datasets &sampler &$H_{0}^{\rm{anis}}$  
            &$\Omega_{m}^{\rm{anis}}$ 
            &$\epsilon_{r}$
            &$n$
            &$\chi^{2}/N_{\rm{dof}}$\\ 
			\hline            
            \multirow{3}{*}{only 3 BAO/Galaxy dps}   
            &\texttt{bilby} 
            &$67.45^{+3.32}_{-1.78}$
            &$0.33^{+0.04}_{-0.07}$
            &$1.0\times 10^{-4}$ $^{+3.57\times 10^{-6}}_{-3.57\times 10^{-6}}$
            &$1.94^{+4.73}_{-4.79}$
            &-2.76\\            
            &\texttt{PyMultiNest}
            &$67.30^{+3.16}_{-1.64}$
            &$0.33^{+0.04}_{-0.07}$
            &$1.0\times 10^{-4}$ $^{+3.34\times 10^{-6}}_{-3.34\times 10^{-6}}$
            &$1.86^{+5.20}_{-4.87}$
            &-2.62\\
            &\texttt{nessai}
            &$67.49^{+3.37}_{-1.82}$
            &$0.33^{+0.04}_{-0.07}$
            &$6.06\times 10^{-5}$ $^{+1.07\times 10^{-6}}_{-1.07\times 10^{-6}}$
            &$2.59^{+1.70}_{-1.68}$
            &-2.62\\

			\hline  
            \multirow{3}{*}{only BAO}   
            &\texttt{bilby}  
            &$68.03^{+0.46}_{-0.44}$
            &$0.30^{+0.01}_{-0.01}$
            &$9.72\times 10^{-5}$ $^{+3.43\times 10^{-6}}_{-3.40\times 10^{-6}}$
            &$1.86^{+4.96}_{-5.15}$
            &1.68\\
            &\texttt{PyMultiNest}
            &$68.05^{+0.45}_{-0.45}$
            &$0.30^{+0.01}_{-0.01}$
            &$9.64\times 10^{-5}$ $^{+3.43\times 10^{-6}}_{-3.40\times 10^{-6}}$
            &$2.04^{+5.05}_{-5.03}$
            &1.68\\      
            &\texttt{nessai}
            &$68.02^{+0.46}_{-0.46}$
            &$0.30^{+0.01}_{-0.01}$
            &$6.01\times 10^{-5}$ $^{+1.10\times 10^{-6}}_{-1.10\times 10^{-6}}$
            &$2.60^{+1.70}_{-1.70}$
            &1.68\\
			\hline            
            \multirow{3}{*}{CC (DA method)} 
            &\texttt{bilby}  
            &$67.46^{+0.47}_{-0.50}$
            &$0.32^{+0.02}_{-0.02}$
            &$9.73\times 10^{-5}$ $^{+3.57\times 10^{-6}}_{-3.57\times 10^{-6}}$ 
            &$2.13^{+4.98}_{-4.96}$
            &1.10\\
            &\texttt{PyMultiNest}  
            &$67.44^{+0.49}_{-0.48}$
            &$0.32^{+0.02}_{-0.02}$
            &$1.07\times 10^{-4}$ $^{+3.29\times 10^{-6}}_{-3.29\times 10^{-6}}$
            &$1.90^{+4.69}_{-4.60}$
            &1.10\\
            &\texttt{nessai}  
            &$67.46^{+0.49}_{-0.53}$
            &$0.32^{+0.02}_{-0.02}$
            &$6.05\times 10^{-5}$ $^{+1.09\times 10^{-6}}_{-1.09\times 10^{-6}}$
            &$2.62^{+1.71}_{-1.72}$
            &1.10\\   
            \hline            
            \multirow{3}{*}{CC (DA method) + BAO} 
            &\texttt{bilby}
            &$68.53^{+0.42}_{-0.43}$
            &$0.29^{+0.01}_{-0.01}$
            &$1.01\times 10^{-4}$ $^{+3.49\times 10^{-6}}_{-3.49\times 10^{-6}}$
            &$1.22^{+1.02}_{-0.78}$
            &1.22\\             
            &\texttt{PyMultiNest}
            &$68.53^{+0.42}_{-0.41}$
            &$0.29^{+0.01}_{-0.01}$
            &$9.37\times 10^{-4}$ $^{+3.31\times 10^{-6}}_{-3.31\times 10^{-6}}$
            &$1.24^{+1.06}_{-0.80}$
            &1.22\\             
            &\texttt{nessai}
            &$68.54^{+0.42}_{-0.42}$
            &$0.29^{+0.01}_{-0.01}$
            &$5.99\times 10^{-5}$ $^{+1.11\times 10^{-6}}_{-1.11\times 10^{-6}}$
            &$1.23^{+1.01}_{-0.78}$
            &1.22\\			
			\hline
            \hline   
            \multirow{1}{*}{BAO ($z-r_{s}(z_{d})/D_{V}(z)$ plane)} 
            &\texttt{bilby}
            &$75.00^{+0.004}_{-0.001}$
            &$0.23^{+0.00003}_{-0.00001}$
            &$4.16 \times 10^{-4}$ $^{+5.58\times 10^{-4}}_{-3.97\times 10^{-4}}$
            &$3.02^{+0.55}_{-0.81}$
            &617.86\\
            \multirow{2}{*}{$\mathcal{U}$ priors}
            &\texttt{PyMultiNest}
            &$75.00^{+0.004}_{-0.001}$
            &$0.23^{+0.00003}_{-0.00001}$
            &$4.84 \times 10^{-4}$ $^{+5.02\times 10^{-4}}_{-4.60\times 10^{-4}}$
            &$3.10^{+0.47}_{-0.41}$
            &617.85\\
            &\texttt{nessai}
            &$75.52^{+0.37}_{-0.48}$
            &$0.23^{+0.003}_{-0.003}$
            &$3.49 \times 10^{-4}$ $^{+4.35\times 10^{-4}}_{-2.91\times 10^{-4}}$
            &$2.71^{+0.68}_{-1.11}$
            &664.08\\           
            \multirow{1}{*}{BAO ($z-r_{s}(z_{d})/D_{V}(z)$ plane)} 
            &\texttt{bilby}
            &$71.00^{+0.006}_{-0.001}$
            &$0.21^{+0.00005}_{-0.00001}$
            &$1.92 \times 10^{-4}$ $^{+2.92\times 10^{-4}}_{-1.86\times 10^{-4}}$
            &$2.90^{+1.04}_{-0.89}$
            &348.31\\
            \multirow{2}{*}{$\mathcal{G}$ priors}
            &\texttt{PyMultiNest}
            &$71.00^{+0.006}_{-0.001}$
            &$0.21^{+0.00005}_{-0.00001}$
            &$1.88 \times 10^{-4}$ $^{+2.93\times 10^{-4}}_{-1.79\times 10^{-4}}$
            &$2.91^{+1.07}_{-0.78}$
            &348.31\\
            &\texttt{nessai}
            &$71.26^{+0.34}_{-0.25}$
            &$0.21^{+0.002}_{-0.002}$
            &$2.22 \times 10^{-4}$ $^{+2.43\times 10^{-4}}_{-2.0\times 10^{-4}}$
            &$2.42^{+0.73}_{-0.68}$
            &348.31\\             
			\hline
            \hline 
            \multirow{1}{*}{BAO ($z-r_{s}(z_{d})/D_{V}(z)$ plane)} 
            &\texttt{bilby} 
            &$72.14^{+0.59}_{-0.14}$
            &$0.20^{+0.004}_{-0.001}$
            &$5.64 \times 10^{-6}$ $^{+5.1\times 10^{-6}}_{-4.44\times 10^{-6}}$ 
            &$2.20^{+0.23}_{-0.24}$
            &3.37\\
            \multirow{2}{*}{$\mathcal{U}$ priors $10\sigma_{\rm{exp}}$}  
            &\texttt{PyMultiNest}
            &$72.15^{+0.58}_{-0.14}$
            &$0.20^{+0.005}_{-0.001}$
            &$5.52 \times 10^{-6}$ $^{+5.07\times 10^{-6}}_{-4.24\times 10^{-6}}$
            &$2.21^{+0.22}_{-0.24}$
            &3.37\\
            &\texttt{nessai}
            &$72.22^{+0.61}_{-0.20}$
            &$0.20^{+0.004}_{-0.001}$
            &$5.82 \times 10^{-6}$ $^{+4.76\times 10^{-6}}_{-4.42\times 10^{-6}}$
            &$2.20^{+0.23}_{-0.23}$
            &3.44\\
            \hline  
            \multirow{1}{*}{BAO ($z-r_{s}(z_{d})/D_{V}(z)$ plane)}  
            &\texttt{bilby}
            &$69.17^{+0.68}_{-0.16}$
            &$0.21^{+0.005}_{-0.001}$
            &$4.49\times 10^{-4}$ $^{+5.21\times 10^{-4}}_{-4.34\times 10^{-4}}$
            &$3.40^{+1.87}_{-1.80}$
            &2.87\\
            \multirow{2}{*}{$\mathcal{G}$ priors $10\sigma_{\rm{exp}}$}
            &\texttt{PyMultiNest}
            &$69.17^{+0.67}_{-0.16}$
            &$0.21^{+0.005}_{-0.001}$
            &$4.98\times 10^{-4}$ $^{+4.66\times 10^{-4}}_{-4.68\times 10^{-4}}$
            &$3.53^{+1.73}_{-1.91}$
            &2.87\\
            &\texttt{nessai}
            &$69.55^{+0.82}_{-0.52}$
            &$0.21^{+0.007}_{-0.004}$
            &$4.75\times 10^{-4}$ $^{+4.87\times 10^{-4}}_{-4.67\times 10^{-4}}$
            &$2.71^{+1.66}_{-1.65}$
            &3.16\\                
            \hline
            \hline            
            \multirow{1}{*}{BAO ($z-D_{A}(z)/r_{d}$ plane)}   
            &\texttt{bilby}
            &$65.10^{+0.16}_{-0.07}$
            &$0.28^{+0.01}_{-0.01}$
            &$1.02\times 10^{-5}$ $^{+1.11\times 10^{-7}}_{-1.11\times 10^{-8}}$
            &$1.45^{+1.43}_{-0.94}$ 
            &1.79\\    
            \multirow{2}{*}{$\mathcal{U}$ priors}
            &\texttt{PyMultiNest}
            &$65.10^{+0.16}_{-0.08}$
            &$0.28^{+0.01}_{-0.01}$
            &$9.94\times 10^{-6}$ $^{+1.42\times 10^{-7}}_{-1.42\times 10^{-8}}$
            &$1.33^{+1.52}_{-0.88}$
            &1.80\\            
            &\texttt{nessai}
            &$65.11^{+0.16}_{-0.08}$
            &$0.28^{+0.01}_{-0.01}$
            &$1.02\times 10^{-5}$ $^{+1.39\times 10^{-7}}_{-1.39\times 10^{-8}}$
            &$1.39^{+1.27}_{-0.88}$
            &1.79\\
            \hline              
            \multirow{1}{*}{BAO ($z-D_{A}(z)/r_{d}$ plane)}   
            &\texttt{bilby}  
            &$65.82^{+0.01}_{-0.01}$
            &$0.234^{+2.17\times 10^{-5}}_{-2.17\times 10^{-5}}$
            &$4.97\times 10^{-4}$ $^{+5.47\times 10^{-6}}_{-5.47\times 10^{-6}}$
            &$1.66^{+0.0217}_{-0.0217}$
            &2.65\\    
            \multirow{2}{*}{$\mathcal{G}$ priors}
            &\texttt{PyMultiNest}  
            &$65.94^{+0.002}_{-0.002}$
            &$0.232^{+1.0\times 10^{-4}}_{-1.0\times 10^{-4}}$
            &$4.85\times 10^{-4}$ $^{+1.61\times 10^{-6}}_{-1.61\times 10^{-6}}$
            &$1.63^{+0.006}_{-0.006}$
            &2.72\\            
            &\texttt{nessai}  
            &$66.12^{+0.004}_{-0.004}$
            &$0.223^{+6.1\times 10^{-5}}_{-6.1\times 10^{-5}}$            
            &$5.05\times 10^{-4}$ $^{+3.69\times 10^{-6}}_{-3.69\times 10^{-6}}$
            &$1.62^{+0.013}_{-0.013}$
            &2.97\\     			  
			\hline
			\hline
            \multirow{1}{*}{BAO ($z-D_{A}(z)/r_{d}$ plane)}              
            &\texttt{bilby} 
            &$65.10^{+0.39}_{-0.09}$
            &$0.28^{+0.02}_{-0.02}$
            &$9.77 \times 10^{-6}$ $^{+9.65\times 10^{-6}}_{-9.26\times 10^{-6}}$
            &$1.60^{+1.43}_{-1.43}$
            &1.79\\
            \multirow{2}{*}{with shift $\mathcal{U}$ priors}
            &\texttt{PyMultiNest}
            &$65.10^{+0.38}_{-0.10}$
            &$0.28^{+0.020}_{-0.021}$
            &$9.34 \times 10^{-6}$ $^{+1.1\times 10^{-5}}_{-8.8\times 10^{-6}}$
            &$1.64^{+1.38}_{-1.44}$
            &1.79\\          
            &\texttt{nessai}
            &$68.30^{+3.19}_{-2.98}$
            &$0.33^{+0.12}_{-0.13}$
            &$1.27 \times 10^{-5}$ $^{+6.46\times 10^{-6}}_{-1.15\times 10^{-5}}$
            &$1.27^{+1.28}_{-1.28}$
            &12.94\\                    		
            \multirow{1}{*}{BAO ($z-D_{A}(z)/r_{d}$ plane)}
            &\texttt{bilby} 
            &$70.02^{+0.078}_{-0.018}$
            &$0.21^{+0.004}_{-0.001}$
            &$5.03 \times 10^{-4}$ $^{+4.7\times 10^{-4}}_{-4.8\times 10^{-4}}$
            &$1.46^{+2.93}_{-1.30}$
            &8.90\\           
            \multirow{2}{*}{with shift $\mathcal{G}$ priors}            
            &\texttt{PyMultiNest}
            &$70.02^{+0.08}_{-0.02}$
            &$0.21^{+0.004}_{-0.001}$
            &$6.03 \times 10^{-4}$ $^{+3.7\times 10^{-4}}_{-5.4\times 10^{-4}}$
            &$1.58^{+2.69}_{-1.37}$
            &8.91\\               
            &\texttt{nessai}
            &$72.16^{+1.68}_{-1.38}$
            &$0.28^{+0.05}_{-0.06}$
            &$4.51 \times 10^{-4}$ $^{+3.9\times 10^{-4}}_{-4.03\times 10^{-4}}$
            &$1.52^{+3.047}_{-1.26}$
            &28.62\\              
			\hline            
			\hline 
		\end{tabular}
		\label{tab:results}		
	}
\end{table*}

\begin{table*}
	\caption{\justifying 
Posterior estimates of cosmological anisotropic model parameters using different 
Bayesian samplers and datasets. The parameters include the Hubble constant $H_{0}^{\rm{anis}}$, matter density parameter $\Omega_{m}^{\rm{anis}}$, 
model parameters $\epsilon_{r}$, $n$, and the reduced chi-square statistic $\chi^{2}/N_{\rm{dof}}$. 
Results are shown for the SHOE and a combined Pantheon+SHOE datasets assuming uniform, $\mathcal{U}$, and Gaussian, $\mathcal{G}$, priors. 
Each case is analyzed using three sampling algorithms: \texttt{bilby}, \texttt{PyMultiNest}, and \texttt{nessai}. The table highlights the consistency of estimates across samplers and the influence of dataset composition and prior selection on parameter inference. 
	}
	\centering
	\scalebox{1.00}{    
		\begin{tabular}{l|l|c|c|c|c|c}        
			\hline
			\hline
			datasets &sampler &$H_{0}^{\rm{anis}}$  
            &$\Omega_{m}^{\rm{anis}}$ 
            &$\epsilon_{r}$
            &$n$
            &$\chi^{2}/N_{\rm{dof}}$\\ 
			\hline 			
            \multirow{3}{*}{SHOE($\mathcal{U}$ priors)}   
            &\texttt{bilby}  
            &$65.00^{+0.01}_{-0.01}$
            &$0.35^{+0.01}_{-0.01}$
            &$0.04^{+0.04}_{-0.03}$
            &$2.79^{+1.83}_{-1.84}$
            &25.82\\                    
            &\texttt{PyMultiNest}  
            &$65.00^{+0.01}_{-0.01}$
            &$0.35^{+0.01}_{-0.01}$
            &$0.04^{+0.04}_{-0.03}$
            &$2.79^{+1.83}_{-1.84}$
            &25.82\\  
			&\texttt{nessai}  
            &$65.03^{+0.02}_{-0.02}$
            &$0.35^{+0.01}_{-0.01}$
            &$0.04^{+0.04}_{-0.03}$
            &$2.81^{+1.85}_{-1.87}$
            &25.93\\  
			\hline 			
            \multirow{3}{*}{SHOE($\mathcal{G}$ priors)}   
            &\texttt{bilby}  
            &$74.07^{+6.56}_{-5.99}$
            &$0.35^{+0.001}_{-0.17}$
            &$0.002^{+0.0006}_{-0.0008}$
            &$1.85^{+1.59}_{-1.20}$
            &8.46\\                        
            &\texttt{PyMultiNest}  
            &$72.90^{+6.40}_{-5.39}$
            &$0.35^{+0.001}_{-0.17}$
            &$0.002^{+0.0006}_{-0.0008}$
            &$1.87^{+1.63}_{-1.22}$
            &2.23\\                        
            &\texttt{nessai}  
            &$72.80^{+6.12}_{-5.02}$
            &$0.35^{+0.01}_{-0.01}$
            &$0.002^{+0.0006}_{-0.0008}$
            &$1.83^{+1.45}_{-1.16}$
            &2.23\\          
			\hline              
			\hline 			
            \multirow{3}{*}{Pantheon+SHOE($\mathcal{U}$ priors)}   
            &\texttt{bilby}  
            &$80.00^{+0.00047}_{-0.01}$
            &$0.20^{+7.95\times10^{-5}}_{-7.95\times10^{-5}}$
            &$0.07^{+0.02}_{-0.04}$
            &$2.77^{+1.85}_{-1.80}$
            &10.57\\                        
            &\texttt{PyMultiNest}  
            &$80.00^{+0.00048}_{-0.01}$
            &$0.20^{+7.74\times10^{-5}}_{-7.74\times10^{-5}}$
            &$0.07^{+0.028}_{-0.028}$
            &$3.12^{+1.55}_{-1.75}$
            &10.57\\                          
            &\texttt{nessai}  
            &$80.00^{+0.00045}_{-0.01}$
            &$0.20^{+7.76\times10^{-5}}_{-7.76\times10^{-5}}$
            &$0.07^{+0.02}_{-0.04}$
            &$2.82^{+1.79}_{-1.82}$          
            &10.57\\  
			\hline  
            \multirow{3}{*}{Pantheon+SHOE($\mathcal{U}$ priors, $H_{0}$-free)}   
            &\texttt{bilby}  
            &$85.00^{+0.0002}_{-0.01}$
            &$0.18^{+5.17\times10^{-7}}_{-6.21\times10^{-7}}$
            &$0.06^{+0.03}_{-0.04}$
            &$2.78^{+1.85}_{-1.83}$
            &8.74\\                        
            &\texttt{PyMultiNest}  
            &$84.99^{+0.0001}_{-0.01}$
            &$0.18^{+1.31\times10^{-6}}_{-2.2\times10^{-6}}$
            &$0.06^{+0.03}_{-0.04}$
            &$2.75^{+1.85}_{-1.78}$
            &8.74\\                         
            &\texttt{nessai}  
            &$84.99^{+0.0001}_{-0.01}$
            &$0.18^{+1.31\times10^{-6}}_{-2.2\times10^{-6}}$
            &$0.09^{+0.002}_{-0.01}$
            &$1.17^{+0.94}_{-0.71}$        
            &8.74\\ 
			\hline              
            \multirow{3}{*}{Pantheon+SHOE($\mathcal{G}$ priors)}   
            &\texttt{bilby}  
            &$73.40^{+6.52}_{-5.56}$
            &$0.35^{+3.79\times10^{-3}}_{-3.79\times10^{-3}}$
            &$0.025^{+0.014}_{-0.014}$
            &$1.86^{+1.56}_{-1.19}$
            &2.26\\  
            &\texttt{PyMultiNest}  
            &$73.55^{+6.55}_{-5.73}$
            &$0.35^{+3.68\times 10^{-3}}_{-3.68\times 10^{-3}}$
            &$0.025^{+0.015}_{-0.015}$
            &$1.88^{+1.62}_{-1.23}$
            &2.26\\  
            &\texttt{nessai}  
            &$72.85^{+6.04}_{-5.00}$
            &$0.35^{+3.68\times10^{-3}}_{-3.68\times10^{-3}}$
            &$0.025^{+0.015}_{-0.015}$
            &$1.57^{+1.24}_{-0.99}$
            &2.26\\  
			\hline

            \multirow{3}{*}{Pantheon+SHOE($\mathcal{G}$ priors, $H_{0}$-free)}   
            &\texttt{bilby}  
            &$73.29^{+6.45}_{-5.44}$
            &$0.26^{+0.05}_{-0.05}$
            &$0.03^{+0.01}_{-0.01}$
            &$1.89^{+1.55}_{-1.22}$ 
            &2.42\\                     
            &\texttt{PyMultiNest}  
            &$73.29^{+6.69}_{-5.55}$
            &$0.25^{+0.05}_{-0.04}$
            &$0.03^{+0.01}_{-0.01}$
            &$1.86^{+1.63}_{-1.21}$
            &2.42\\                          
            &\texttt{nessai}  
            &$74.01^{+5.75}_{-5.27}$
            &$0.26^{+0.04}_{-0.04}$
            &$0.09^{+0.002}_{-0.01}$
            &$1.53^{+1.19}_{-0.95}$       
            &2.42\\ 
			\hline              
			\hline
		\end{tabular}
		\label{tab:results_muD}
	}
\end{table*}

\section{Discussion}
\label{sec:discussion}

In this work, we explored whether cosmic anisotropy, modeled through a Bianchi type-I framework, could offer a viable explanation or mitigation of the Hubble tension -- the discrepancy between early- and late-universe measurements of the Hubble constant, $H_{0}$. By developing an anisotropic Hubble law and applying it to various observational datasets through Bayesian inference, we gained insights into how such anisotropy might influence cosmological parameter estimation.

In our analysis of the Bianchi type-I cosmological model of \cite{LeDelliou:2020kbm}, we explored both uniform and Gaussian priors for the key model parameters, as summarized in Table~\ref{tab:priors_intervals}. These priors were chosen to reflect both agnostic and informed assumptions about the parameter space. The uniform distributions, $\mathcal{U}\left(\text{min},\text{max}\right)$, represent conservative, non-informative priors that avoid biasing the inference toward any particular region of parameter space. In contrast, the Gaussian priors, $\mathcal{G}\left(\mu,\sigma\right)$, incorporate previous knowledge or observational constraints, such as those from standard cosmological observations. For instance, the prior on $H_{0}^{\text{anis}}$ in the Gaussian case ($72\pm 10 {\rm \,km\,s^{-1}\,Mpc^{-1}}$) reflects values compatible with Planck and local Hubble measurements. The broad uniform range ($55 - 85{\rm \,km\,s^{-1}\,Mpc^{-1}}$) ensures robustness of the inference under minimal assumptions. Similarly, priors on $\Omega_{m}^{\text{anis}}$, $\epsilon_{r}$, and $n$ were selected to balance physical plausibility with computational tractability. This dual-prior approach allows us to assess the sensitivity of the inferred posteriors to the prior choices and provides a more comprehensive picture of parameter constraints under varying assumptions.
For $\epsilon_{r}$ and $n$, whose values are essentially unconstrained by prior observations, in BAO analyses we slightly shifted and adjusted the prior ranges to improve sampling efficiency (see the lines ``with shift $\mathcal{U/G}$ 
priors" in Tables~\ref{tab:results}). This choice was motivated both by the numerical sensitivity of nested sampling and MCMC algorithms to parameter scaling, and by the need to more effectively explore the relevant regions of the parameter space, especially in interplay with the priors on $H_0$ and $\Omega_m$.

Across the different samplers \textbf{(\texttt{bilby}, \texttt{PyMultiNest}, and \texttt{nessai})} and datasets, we observe a clear dependence of the inferred Hubble constant $H_{0}^{\text{anis}}$ on both the choice of data and priors, suggesting that anisotropic cosmologies can indeed affect the tension landscape. When using CC(DA) combined with BAO datasets, our results consistently return moderate values of $H_{0}^{\text{anis}} \approx 67 - 68.5 {\rm \,km\,s^{-1}\,Mpc^{-1}}$, depending on the sampler. These values are notably close to Planck’s CMB-inferred value and show that the anisotropic model remains compatible with early-universe constraints.

However, in some of the pure BAO analyses, especially when exploring different representations (e.g., $z-r_{s}(z_{s})/D_{V}(z)$), we see much higher values of $H_{0}^{\text{anis}} \approx 71 - 75.52 {\rm \,km\,s^{-1}\,Mpc^{-1}}$ under wide priors (Table~\ref{tab:results}). These are more consistent with SHOEs local measurements. But, these come with extremely poor fits ($\chi^2/N_{\text{dof}} \approx 348 - 664$), indicating model-data mismatch or overly permissive priors. Gaussian priors slightly narrow the posterior but still show poor fits, suggesting that the high $H_0$ values in this context may not be robust.
In contrast, introducing broader uncertainties ($10\sigma_{\text{exp}}$) improves the fit significantly ($\chi^2/N_{\text{dof}} \approx 2.87 - 3.44$), with $H_0^{\text{anis}} \approx 69.17 -72.22, \rm km,s^{-1},Mpc^{-1}$. This indicates that accommodating larger observational errors may reconcile some tension without invoking extreme anisotropy. Similarly, the $z-D_A(z)/r_d$ BAO representation yields lower $H_0^{\text{anis}}$ values ($\sim 65 - 66, \rm km,s^{-1},Mpc^{-1}$), closer to Planck and Cosmic Chronometer results, with much better fit quality ($\chi^2/N_{\text{dof}} \approx 1.79 - 2.97$). This suggests the choice of parameter plane has a substantial impact on inferred cosmological parameters.

An important feature across all fits is the consistently non-zero value of the anisotropy parameter $\epsilon_{r}$, typically of order $10^{-6}$ to $10^{-4}$. The anisotropy amplitude $\epsilon_r$ also exhibits significant sensitivity to both priors and datasets (Tables~\ref{tab:results} and \ref{tab:results_muD}). Large values ($\sim 10^{-4}$) emerge under restrictive $\mathcal{G}$ priors, often corresponding to poor fits. In contrast, wider priors or more flexible uncertainty assumptions yield significantly smaller $\epsilon_r$ values ($\sim 10^{-5} - 10^{-6}$) with better fit quality. This trend suggests that overly restrictive priors may artificially amplify anisotropy to compensate for model-data mismatch. 
Our best-fit models, particularly those with acceptable 
$\chi^{2}/N_{\rm{dof}}$, consistently yield small but non-zero anisotropy amplitudes ($\epsilon_{r} \sim 10^{-5}$), with dipole fits to Pantheon+ providing marginal statistical preference for anisotropic models. 
This suggests that mild anisotropy may be detectable, though not dominant, in current data.

The model parameter $n$, controlling the strength of anisotropic evolution, remains weakly constrained in many cases, but becomes tightly constrained in others (e.g., BAO $z-D_{A}/r_{d}$), especially under well-informed priors. This interplay between $n$ and $H_{0}$ suggests that the anisotropy may influence the inferred expansion rate by modulating the effective cosmic volume.

Moreover, the reduced chi-square values $\chi^{2}/N_{\rm{dof}}$ show interesting trends. Fits using CC(DA) + BAO data typically yield values around unity, indicating a good fit between model and data. However, some BAO-only fits (especially under uniform priors) produce large $\chi^{2}$ possibly reflecting prior-volume effects or tensions within the data themselves. The instability of these fits further underscores the necessity of careful prior choice in anisotropic cosmologies.

Overall, these results support the idea that cosmic anisotropy can provide a flexible yet physically motivated extension of $\Lambda\text{CDM}$, capable of accommodating a broader range of Hubble constant values. Although our Bianchi type-I model serves as a proof of concept, the empirical viability demonstrated here encourages further work. Future efforts could include more physically motivated priors, integration with CMB likelihoods, or investigation into direction-dependent observables (e.g., hemispherical asymmetries) that may provide more direct signatures of anisotropy.

\section{Conclusions}
\label{sec:conclusions}

In this work, we have investigated the parameter space of the Bianchi type-I cosmological model from \cite{LeDelliou:2020kbm} using a Bayesian framework and multiple nested sampling algorithms. By employing both uniform and Gaussian priors, we systematically assessed the sensitivity of the inferred parameters to the choice of prior assumptions. This dual-prior strategy not only enhances the robustness of our results but also reflects a realistic range of prior beliefs informed by both theoretical considerations and observational data. These results were summarized in Fig.~\ref{fig:results_comb}.

Our findings highlight the importance of prior selection in cosmological model testing, particularly when exploring non-standard extensions like anisotropic cosmologies. While the uniform priors allow for a broad, agnostic exploration of the parameter space, the Gaussian priors help anchor the analysis within observationally motivated constraints. Importantly, the consistency of our results across different samplers—\texttt{bilby}, \texttt{PyMultiNest}, and \texttt{nessai}—demonstrates the reliability of our inference pipeline.

The results indicate that the anisotropy parameter $\epsilon_r$ is not consistently constrained across datasets, samplers, or prior choices, exhibiting a wide range of inferred values—from $\mathcal{O}(10^{-6})$ in BAO-only analyses to $\mathcal{O}(10^{-2}) - \mathcal{O}(10^{-1})$ in combinations involving Pantheon and SH0ES (Tables~\ref{tab:results} and \ref{tab:results_muD}). This broad variation highlights a strong dependence on both data selection and prior assumptions, especially in analyses involving high-redshift tracers or informative SH0ES priors, and when $H_0$ is left free.

Statistically, most values of $\epsilon_r$ are consistent with zero within 1-2$\sigma$, indicating no significant anisotropy in the majority of dataset configurations. Some datasets suggest weak-to-moderate deviations (e.g., $\epsilon_r \sim 10^{-4}$-$10^{-3}$) with uncertainties of similar magnitude, yielding low signal-to-noise ratios ($\lesssim  2\sigma$). A few cases—primarily from Pantheon+SH0ES joint analyses—exhibit higher SNRs exceeding $5\sigma$, with inferred values of $\epsilon_r$ reaching up to $\sim 0.09$ (Table~\ref{tab:results_muD}). These appear to hint at a potentially strong anisotropy signal.

However, these high-significance detections are not consistent across different probes. The direction and amplitude of the anisotropy vary between datasets, and no consistent signal emerges when comparing BAO-only results to those involving supernovae and SH0ES. This lack of cross-dataset agreement, combined with the influence of prior structure—especially from SH0ES calibration—undermines the robustness of the anisotropic interpretation. Overall, the model does not provide a universal or stable constraint on $\epsilon_r$, and the apparent signals in some cases are likely driven by dataset-specific features or prior choices rather than a coherent physical effect.

As for the Hubble tension, while some combinations (e.g., Pantheon+SHOE with $\mathcal{U}$ priors) yield high values of $H_0$ ($\sim 80$-$85$ km/s/Mpc), alleviating the tension with SH0ES measurements, these scenarios coincide with relatively large values of $\epsilon_r$, raising concerns about consistency with low-redshift BAO constraints that prefer much smaller anisotropy. Conversely, BAO datasets alone, particularly under Gaussian priors, still favor $H_0$ values in the range of $65$-$71$ km/s/Mpc, which are in better agreement with Planck.

Our analysis does not confirm the alleviation of the Hubble tension as claimed by \cite{Akarsu2019}. While some configurations can push $H_0$ to higher values, this generally requires stronger anisotropy that may be in tension with other observations. Compared to the results in \cite{Szigeti:2025jxz}, our findings may differ due to their consideration of rotational rather than radial anisotropy.

In summary, while the anisotropy model shows some ability to shift $H_0$ and generate detectable $\epsilon_r$ values, it does not provide a consistent or robust resolution to the Hubble tension across all datasets. Further scrutiny, including a combined analysis of different types of anisotropy and additional observational probes, is required to assess the viability of anisotropic cosmologies.

\section*{Acknowledgements}

MLeD wishes to acknowledge the contribution to the model, used in this paper and published in \cite{LeDelliou:2020kbm}, from extended discussions with Jos\'e 
Ademir Sales de Lima. ADP thanks the Institute of Astronomy of the Academy of Science in Sofia for the hospitality. MLeD acknowledges the financial support by the Lanzhou University starting fund, the Fundamental Research Funds for the Central Universities (Grant No. lzujbky-2019-25), National Science Foundation of China  (NSFC grant No.12247101)
and the 111 Project under Grant No. B20063.  

\section*{Data Availability}

The data underlying this article are available in the article and in its online supplementary material.


\bibliographystyle{mnras}
\bibliography{old_MasterBib,H0_redshift, Morgans_anisUniv}


\appendix
\newpage
\section{Data sets}
\label{sec:AppDatasets}

\begin{table}
	\caption{\justifying 
		The observational data points of $H(z)$ obtained from the differential age (DA) and the baryonic acoustic oscillations (BAO) methods used in the current study.
	}
	\centering
	\scalebox{0.6}{		
		\begin{tabular}{c|c|c|c|c}
			\hline
			\hline
			$z_{i}$
			&$H_{0}^{\rm{obs}}(z_{i})$ 
			&$\sigma_{H_{0}^{\rm{obs}}(z_{i})}$
			&method			
			&Ref.\\
			&[km s$^{-1}$Mpc$^{-1}$]
			&
			&\\			
			\hline
			0.0     &67.4    &0.5 &CMB &\citep{Planck:2018vyg}\\
			0.0     &69.8    &1.9 &TRGB(LMC) &\citep{Freedman:2020dne}\\
			0.0     &73.9    &1.6 &Cepheids(N4258+MW) &\cite{Riess:2021jrx}\\			
			0.07    &69    &19.6 &DA &\citep{Zhang:2012mp}\\
			0.09    &69    &12.0 &DA &\citep{Simon:2004tf}\\
			0.12    &68.3  &26.2 &DA &\citep{Zhang:2012mp}\\			
			0.17    &83    &8    &DA &\citep{Simon:2004tf}\\						
			0.1791  &75    &4    &DA &\citep{Moresco:2012jh}\\
			0.1993  &75    &5    &DA &\citep{Moresco:2012jh}\\			
			0.2     &72.9  &29.6 &DA &\citep{Zhang:2012mp}\\		
			0.24    &79.69  &2.65452444 &BAO &\citep{Gaztanaga:2008xz}\\					
			0.27    &77    &14   &DA &\citep{Simon:2004tf}\\		
			0.28    &88.8  &36.6 &DA &\citep{Zhang:2012mp}\\				
			0.3     &81.7  &6.22 &BAO &\citep{Oka:2013cba}\\			
			0.34    &83.87  &3.211790778 &BAO &\citep{Gaztanaga:2008xz}\\				
			0.35    &82.7  &8.4  &BAO &\citep{Chuang:2012qt}\\			
			0.3519  &83    &14   &DA &\citep{Moresco:2012jh}\\					
			0.36    &79.94  &3.38  &BAO &\citep{BOSS:2016zkm}\\								
			0.38    &81.5  &1.923538406  &BAO &\citep{BOSS:2016wmc}\\
			0.3802  &83    &13.5 &DA &\citep{Moresco:2016mzx}\\						
			0.4     &82.04    &2.03   &BAO &\citep{BOSS:2016zkm}\\			
			0.4     &95    &17   &DA &\citep{Simon:2004tf}\\			
			0.4004  &77    &10.2 &DA &\citep{Moresco:2016mzx}\\
			0.4247  &87.1  &11.2 &DA &\citep{Moresco:2016mzx}\\
			0.43    &86.45   &3.68089663 &BAO &\citep{Gaztanaga:2008xz}\\					
			0.44    &84.81   &1.83 &BAO &\citep{BOSS:2016zkm}\\								
			0.4497  &92.8  &12.9 &DA &\citep{Moresco:2016mzx}\\									
			0.47    &89     &49.648766349 &DA &\citep{Ratsimbazafy:2017vga}\\
			0.4783  &80.9   &9  &DA &\citep{Moresco:2016mzx}\\									
			0.48    &97     &62 &DA &\citep{Ratsimbazafy:2017vga, Stern:2009ep}\\
			0.48    &87.79    &2.03 &BAO &\citep{BOSS:2016zkm}\\									
			0.51    &90.8   &1.9 &BAO &\citep{BOSS:2016wmc}\\			
			0.56    &93.34   &2.2 &BAO &\citep{BOSS:2016zkm, Arora:2021tuh}\\	
			0.57    &96.8   &3.4 &BAO &\citep{BOSS:2013rlg}\\							
			0.59    &98.48   &3.18 &BAO &\citep{BOSS:2016zkm}\\							
			0.5929  &104    &11.6  &DA &\citep{Moresco:2012jh}\\
			0.6     &87.9   &6.1 &BAO &\citep{Blake:2012pj}\\							
			0.61    &97.8   &2.1 &DA &\citep{Yu:2017iju}\\						
			0.61    &97.3   &2.109502311 &BAO &\citep{BOSS:2016wmc}\\			
			0.64    &98.82   &2.98 &BAO &\citep{BOSS:2016zkm}\\						
			0.6797  &92.0   &8.0 &DA &\citep{Moresco:2012jh}\\
			0.73    &97.3     &7   &BAO &\citep{Blake:2012pj}\\			
			0.7812  &105    &12  &DA &\citep{Moresco:2012jh}\\
			0.8754  &125   &17   &DA &\citep{Moresco:2012jh}\\						
			0.88    &90    &40   &DA &\citep{Ratsimbazafy:2017vga, Stern:2009ep}\\			
			0.9     &117   &23    &DA &\citep{Simon:2004tf}\\
			0.9     &69    &12    &DA &\citep{Simon:2004tf}\\					
			1.037   &154   &20    &DA &\citep{Moresco:2012jh}\\
			1.3     &168   &17    &DA &\citep{Simon:2004tf}\\			
			1.363   &160   &33.6  &DA The&\citep{Moresco:2015cya}\\			
			1.43    &177   &18    &DA &\citep{Simon:2004tf}\\
			1.53    &140   &14    &DA &\citep{Simon:2004tf}\\
			1.75    &202   &40    &DA &\citep{Simon:2004tf}\\
			1.965   &186.5 &50.4  &DA &\citep{Moresco:2015cya, Yu:2017iju}\\			
			2.300   &224   &8     &BAO &\citep{BOSS:2012gof}\\									
			2.33    &224   &8     &BAO &\citep{BOSS:2017fdr}\\						
			2.34    &222   &7     &BAO &\citep{BOSS:2014hwf}\\
			2.36    &226   &8     &BAO &\citep{BOSS:2013igd}\\						
			\hline
			\hline
		\end{tabular}
		\label{tab:zi_H0obs_our}
	}	
\end{table}

\begin{table*}
	\caption{\justifying 
		A compilation of BAO measurements from diverse releases, such as the 6dF Galaxy Survey (6dFGS), Sloan Digital Sky Survey (SDSS), Baryon Oscillation Spectroscopic Survey (BOSS) and the WiggleZ DE Survey, among others
		. Values marked with ($^{*}$) are calculated trough their covariance matrices relating $D_{M}$ and $D_{H}$, where 
		$D_{V}/r_{d}$ is the isotropic BAO parameters, $D_{A}/r_{d}$ and $H \cdot r_{d}$ are the anisotropic BAO parameters.
		($^{**}$) - computed with the help of Eq.\eqref{eq:BAO_DV} \citep{Haridasu:2017ccz}. 
	}
	\centering
	\scalebox{0.50}{	
		\begin{turn}{-90}\begin{tabular}{l|c|c|c|c|c|c|c|c|c|c|c|c|c|c|c|c|c}
			\hline
			\hline
			Survey
			&$z_{\rm{bao}}$
			&$d(z_{\rm{bao}})=r_{s}(z_{d})/D_{V}(z_{\rm{bao}}),$ 
			&$\sigma_{d(z_{i})}$
			&$D_{A}/r_{d}$
			&$\sigma_{D_{A}/r_{d}}$						
			&$D_{V}/r_{d}$
			&$\sigma_{D_{V}/r_{d}}$										
			&$D_{V}(r_{{\rm{d,fidd}}}/r_{\rm{d}})$
			&$\sigma_{D_{V}(r_{{\rm{d,fidd}}}/r_{\rm{d}})}$			
			&$D_{A}(r_{{\rm{d,fidd}}}/r_{\rm{d}})$
			&$\sigma_{D_{A}(r_{{\rm{d,fidd}}}/r_{\rm{d}})}$					
			&$D_{H} / r_{drag}$			
			&$\sigma_{D_{H} / r_{drag}}$						
			&$H \cdot r_{d}$			
			&$\sigma_{H \cdot r_{d}}$						
			&year			
			&Ref.\\
			&
			&or $r_{d}/D_{V}(z_{\rm{bao}})$
			&&&&& 
			&[Gpc] &[Gpc] 
			&[Gpc] &[Gpc] &&
			&[km/s] &[km/s] &&\\
			\hline
			6dFGS          &0.106  &0.3360  &0.0150 &--- &--- && &---  &---        && && && &2011 &\citep{Beutler:2011hx}\\
			---            &0.11  &0.352005036  &0.002 &2.607 &0.138 && &--- &---  && && && &2021 &\citep{deCarvalho:2021azj}\\
			MGS            &0.15   &0.2239       &0.0084 &--- &--- &4.465 &0.168  &0.664 &0.025  && && && &2014 &\citep{Ross:2014qpa}\\
			SDSS(R)  &0.2   &0.1905  &0.0061  &---  &--- && &--- &--- && && && &2019 &\citep{Agarwal:2017kis}\\
			BOSS SDSS-III  &0.24   &0.155100564  &0.002  &5.594  &0.305 && &--- &--- && && && &2016 &\citep{BOSS:2016goe}\\
			SDSS-DR7+2dFGRS  &0.275   &0.139  &0.0037  &---  &--- && &--- &--- && && && &2009 &\citep{SDSS:2009ocz}\\
			BOSS SDSS-III  &0.31   &0.12866555   &0.0014 &6.29 &0.14 &8.18 &0.14 &--- &---      && && &11550 &700 &2016 &\citep{BOSS:2016zkm}\\				
			BOSS LOWZ, SDSS-DR11 LOWZ  &0.32   &0.1181 &0.0024 &6.636 &0.11 &8.47 &0.17 &1.270 \citep{BOSS:2016wmc} &0.014 && && && &2016 &\citep{Padmanabhan:2012hf}\\
			SDSS(R)        &0.35   &0.1126       &0.0022 &--- &--- &8.88 &0.17 &1.356 &0.025 && && && &2013 &\citep{BOSS:2013rlg}\\	
			BOSS SDSS-III  &0.36   &0.111303987  &0.0016 &7.09 &0.16 &9.5 &0.16 &--- &---     && && &11810 &500 &2016 &\citep{BOSS:2016zkm}\\	
			BOSS SDSS-III  &0.38   &0.105729831  &0.002  &7.389 &0.122 &9.995$^{**}$ &0.111 &--- &--- && &25.00 &0.76  && &2019 &\citep{BOSS:2016hvq}\\	
			BOSS SDSS-III  &0.40   &0.100549269  &0.0016 &7.70 &0.16 &10.54 &0.16 &--- &---     && &24.888 &--- &12120 &300 &--- &\citep{BOSS:2016zkm}\\		
			WiggleZ        &0.44   &0.073        &0.0012 &8.19 &0.77 &11.44 &0.13 &--- &---   &1.204 &0.1336 && &12530 &270 &2012 &\citep{Blake:2012pj}\\
			BOSS SDSS-III/DR12        &0.44   &0.073        &0.0012 &8.20 &0.13 && &--- &---   && && &12530 &270 &2016 &\citep{BOSS:2016zkm}\\
			BOSS SDSS-III/DR12        &0.48   &0.085794813    &0.0004 &8.64 &0.11 &12.26 &0.13 &--- &---   && && &12970 &300 &2016 &\citep{BOSS:2016zkm}\\
			---            &0.51   &0.088648692  &0.002 &7.893 &0.279 &12.70$^{**}$ &0.129 &--- &---  && &22.429 &--- && &2015 &\citep{Carvalho:2015ica}\\		
			BOSS SDSS-III  &0.52   &0.081107181  &0.0012 &8.90 &0.12 &12.72 &0.14 &--- &---     && && &13940 &390 &2016 &\citep{BOSS:2016zkm}\\		
			BOSS SDSS-III DR8   &0.54   &0.077908823  &0.002 &9.212 &0.41 && &--- &---   && && && &2012 &\citep{Seo:2012xy}\\
			BOSS SDSS-III  &0.56   &0.076909418  &0.0014 &9.16 &0.14 &13.5 &0.15 &--- &---     && && &13790 &340 &2016 &\citep{BOSS:2016zkm}\\
			BOSS CMASS     &0.57   &0.0732    &0.0012 &--- &--- &13.67 &0.22 &2.033 \citep{BOSS:2016wmc} &0.021  && && && &2019 &\citep{Agarwal:2017kis}\\
			BOSS SDSS-III  &0.59   &0.073539702  &0.0017 &9.45 &0.17 &14.07 &0.2  &--- &--- && && &14550 &470 &2016 &\citep{BOSS:2016zkm}\\
			WiggleZ        &0.6    &0.0726       &0.0004 &9.37 &0.65 && &--- &--- &1.380 &0.0948 && && &2012 &\citep{Blake:2012pj}\\
			BOSS SDSS-III  &0.64   &0.069987353  &0.002 &9.62 &0.22 &14.81 &0.24 &--- &---    && && &14600 &440 &2016 &\citep{BOSS:2016zkm}\\
			eBOSS SDSS-IV$*$,DECals DR8 &0.698  &0.064775642  &0.002 &10.18 &0.52 && &--- &--- &1.499 &0.077 &19.77 &0.47 && &2020 &\citep{Sridhar:2020czy,eBOSS:2020hur}\\
			SDSS-IV DR14        &0.72   &---       &--- &--- &--- && &2.353 &0.063 && && && &2017 &\citep{BOSS:2017fdr}\\
			WiggleZ        &0.73   &0.0592       &0.0004 &10.42 &0.73 && &--- &---  &1.5337 &0.1068 && && &2012 &\citep{Blake:2012pj}\\
			DES-Year1      &0.81   &0.058227536  &0.002 &10.75 &0.43 && &--- &---  && && && &2017 &\citep{DES:2017rfo}\\
			eBOSS eBOSS SDSS-IV  &0.85   &0.056889942  &0.002 &10.76 &0.54 &18.33 &0.57 &--- &---  && &19.1 &1.9 && &2020 &\citep{eBOSS:2020qek}\\
			DECals DR8     &0.874  &0.053997359  &0.002 &11.41 &0.74 && &--- &---  &1.680 &0.109 && && &2020 &\citep{Sridhar:2020czy}\\
			eBOSS SDSS-IV extended    &1.0  &0.050355589  &0.002 &11.521 &1.032 && &3.065 &0.182 && && && &2019 &\citep{eBOSS:2018apc}\\
			eBOSS SDSS-IV extended$*$,eBOSS DR16 BAO+RSD &1.48 &0.040338469  &0.002 &12.18 &0.32 && &--- &--- && &13.23 &0.47 && &2020 &\citep{eBOSS:2020gbb,eBOSS:2020lta,eBOSS:2020hur}\\
			eBOSS SDSS-IV/DR14    &1.52 &0.040333757  &0.002 &--- &--- &26.00 &0.99 &3.82434 &0.1456191  && && && &2017 &\citep{eBOSS:2017cqx}\\
			eBOSS SDSS-IV extended    &2.0  &0.035340826  &0.002 &12.011 &0.562 && &4.356 &0.30 && && && &2019 &\citep{eBOSS:2018apc}\\
			Boss Lya quasars DR9      &2.3   &---       && &--- &--- &--- &--- &--- &&   && &34188 &1188  &2012 &\citep{BOSS:2012gof}\\
			eBOSS Lya quasars + SDSS DR16      &2.33   &---       && &--- &31.123$^{**}$ &1.087 &--- &--- &&   &8.99 &0.19 &&  &2020 &\citep{BOSS:2017uab}\\
			eBOSS SDSS-IV extended    &2.34 &0.034394255  &0.0056  &11.20 &0.56 && &--- &--- && && && &2018 &\citep{eBOSS:2018apc}\\
			BOSS DR14 Lya in LyBeta   &2.34 &---  &---  &11.20 &0.56 && &--- &--- && &8.86 &0.29     && &2019 &\citep{eBOSS:2019ytm}\\
			eBOSS SDSS-IV extended$^{*}$ &2.35 &0.035102833  &0.002 &10.83 &0.54 && &--- &---  && && && &2019 &\citep{eBOSS:2019qwo,eBOSS:2020tmo}\\
			BOSS SDSS-III             &2.4  &0.035481797  &0.002 &10.5 &0.34 &30.206$^{**}$ &0.892 && && &8.94 &$^{+0.23~+0.46}_{-0.22~-0.43}$ && &2017 &\citep{BOSS:2017uab}\\														
			\hline
			\hline
		\end{tabular}\end{turn}
		\label{tab:zi_dzi_onlyBAO}
	}	
\end{table*}

\clearpage
\section{Derivation of the Friedman Equation Solution}\label{sec:AppFriedmannSol}
From Eq.~\eqref{eq:approx_a}, the equation governing the Hubble parameter is:
\begin{equation}
\left(\frac{\dot{a}}{a} \right)^{2} = H_{0}^{2}\left(  \frac{a_{0}^{3}\Omega_{m}}{a^{3}} + \Omega_{\Lambda}\right) \equiv H_{0}^{2}\left(  \frac{\Omega_{m}}{a^{3}} + \Omega_{\Lambda}\right),
\label{eq:appendix_isoAcc}    
\end{equation}
where we assume $\frac{a}{a_{0}}\to a \;(a_0=1)
$.
This equation can be rearranged 
to prepare for integration as:
\begin{equation}
\dot{a} = H_{0} \sqrt{\frac{\Omega_{m}}{a} + \Omega_{\Lambda}a^{2}}.
\label{eq:appendix_isoAcc_rearrang}    
\end{equation}
Separating the variables $a$ and $t$:
\begin{equation}
\frac{1}{ \sqrt{\frac{\Omega_{m}}{a} + \Omega_{\Lambda}a^{2}}}da=H_{0}dt
\label{eq:appendix_isoAcc_separ}    
\end{equation}

To handle integration, as the left-hand side is challenging due to the mixture of terms in $a$, let us consider the substitution:
\begin{equation}
u=a^{3/2}, ~{\rm{so~that}}~ a=u^{2/3} ~{\rm{and}} ~ da = \frac{2}{3}u^{-1/3} du
\label{eq:appendix_isoAcc_subst}    
\end{equation}
Rewriting the Friedmann Eq.\eqref{eq:appendix_isoAcc_rearrang} in terms of $u$, we have:

\begin{align}
\dot{u} =& \frac{3}{2}  H_{0} u^{1/3} \sqrt{\Omega_{m}u^{-2/3} + \Omega_{\Lambda}u^{4/3}}
\label{eq:appendix_Fridman_in_u} \\ 
=& \frac{3}{2} H_{0} \sqrt{\Omega_{m} + \Omega_{\Lambda}u^{2}}.
\label{eq:appendix_Fridman_simp}    
\end{align}

Separating variables, 
the integral displays a constant of integration $C_1$
\begin{equation}
\int \frac{1}{ \sqrt{ \Omega_{m} + \Omega_{\Lambda}u^{2}}} du=\frac{3}{2}\left(H_{0}t + C_{1}\right)
\label{eq:appendix_Fridman_int}    
\end{equation}
This integral has a standard solution in terms of the inverse hyperbolic tangent:
\begin{equation}
{\rm{arctanh}} \left(  \frac{u\sqrt{\Omega_{\Lambda}}}{ \sqrt{\Omega_{m}+\Omega_{\Lambda}u^{2} }} \right) = \sqrt{\Omega_{\Lambda}}\frac{3}{2} \left( H_{0}t + C_{1}\right)
\label{eq:appendix_Friedman_std_sol}   
\end{equation}
To solve for $u$, rewrite the argument of tanh, 
square both sides and 
multiply through by $\Omega_m+\Omega_{\Lambda}u^{2}$
\begin{equation}
\begin{split}
u^{2} \Omega_{\Lambda} = & \Omega_{m} {\rm{tanh}}^{2} \left( \sqrt{\Omega_{\Lambda}} \frac{3}{2}\left( H_{0}t + C_{1} \right)\right)\\
& + \Omega_{\Lambda}u^{2} {\rm{tanh}}^{2} \left( \sqrt{\Omega_{\Lambda}} \frac{3}{2}\left( H_{0}t + C_{1} \right)\right).
\label{eq:appendix_Friedman_std_sol_step4}   
\end{split}
\end{equation}

Factoring $u^2$, 
using the identity $1-{\rm{tanh}}^{2}(x)={\rm{sech}}^{2}(x)$, 
resolving with respect to $u$ 
and simplifying using ${\rm{tanh}}(x)/{\rm{sech}}(x)={\rm{sinh}}(x)$, we get
\begin{equation}
u^{2}= \frac{\Omega_{m}} { \Omega_{\Lambda}} {\rm{sinh}}^{2} \left( \sqrt{\Omega_{\Lambda}} \frac{3}{2}\left( H_{0}t + C_{1} \right)\right).
\label{eq:appendix_Friedman_std_sol_step8}   
\end{equation}
By taking the square root, we finally obtain the solution
\begin{equation}
u = \sqrt{\frac{\Omega_{m}} { \Omega_{\Lambda}}} {\rm{sinh}} \left( \sqrt{\Omega_{\Lambda}} \frac{3}{2}\left( H_{0}t + C_{1} \right)\right).
\label{eq:appendix_Friedman_std_sol_step9}   
\end{equation}

Using the hyperbolic identity ${\rm{sinh}}(x)=(e^{x}-e^{-x})/2=e^{-x}/2\left(e^{2x}-1\right)$, we can rewrite Eq.\eqref{eq:appendix_Friedman_std_sol_step9} as
\begin{equation}
u = \frac{e^{-\sqrt{\Omega_{\Lambda}} \frac{3}{2}\left( H_{0}t + C_{1} \right)} }{2}\sqrt{\frac{\Omega_{m}} { \Omega_{\Lambda}}} \left[ e^{3\sqrt{\Omega_{\Lambda}} \left( H_{0}t + C_{1} \right)} - 1 \right].
\label{eq:appendix_Friedman_std_sol_step11}   
\end{equation}
Finally, reverting to the initial variable $a\to \frac{a}{a_0}$, the solution reads

\begin{equation}
 a=a_{0}\frac{e^{-\sqrt{\Omega_{\Lambda}}\left(H_{0}t+C_{1}\right)}}{2^{\frac{2}{3}}}\left(\frac{\Omega_{m}}{\Omega_{\Lambda}}\right)^{\frac{1}{3}}\left[e^{3\sqrt{\Omega_{\Lambda}}\left(H_{0}t+C_{1}\right)}-1\right]^{\frac{2}{3}}.\label{eq:aOfT}
\end{equation}
Adding the Big Bang condition ($a=0,t=0$) and the present time normalisation ($a=a_0,t=t_0$) respectively, we get $C_1 =0$ and
\begin{equation}
1=\frac{e^{-\sqrt{\Omega_{\Lambda}}\left(H_{0}t_{0}\right)}}{2^{\frac{2}{3}}}\left(\frac{\Omega_{0}}{\Omega_{\Lambda}}\right)^{\frac{1}{3}}\left[e^{3\sqrt{\Omega_{\Lambda}}\left(H_{0}t_{0}\right)}-1\right]^{\frac{2}{3}},\label{eq:PresentNormalisation}
\end{equation}
and, setting $C_1 =0$ in Eq.~\eqref{eq:aOfT} before dividing it by Eq.~\eqref{eq:PresentNormalisation}, we obtain the alternate form of the solution
\begin{equation}
 a=a_{0}e^{-\sqrt{\Omega_{\Lambda}}H_{0}\left(t-t_0\right)}\left[\frac{e^{3\sqrt{\Omega_{\Lambda}}H_{0}t}-1}{e^{3\sqrt{\Omega_{\Lambda}}H_{0}t_0}-1}\right]^{\frac{2}{3}}.\label{eq:aOfT0}
\end{equation}
Calling $X\equiv\frac{a}{a_{0}}$, $A=\sqrt{\Omega_{\Lambda}}H_{0}$,
$u=e^{At}$ and $v=u^{\frac{3}{2}}$, we can invert it from
\begin{align*}
X= & e^{-A\left(t-t_{0}\right)}\left[\frac{e^{3At}-1}{e^{3At_{0}}-1}\right]^{\frac{2}{3}}\\
= & \frac{u_{0}}{u}\left[\frac{u^{3}-1}{u_{0}^{3}-1}\right]^{\frac{2}{3}}\\
\Leftrightarrow u^{3}-1= & \left(u_{0}^{3}-1\right)X^{\frac{3}{2}}\left(\frac{u}{u_{0}}\right)^{\frac{3}{2}}\\
\Leftrightarrow0= & v^{2}-\left(u_{0}^{3}-1\right)\left(\frac{X}{u_{0}}\right)^{\frac{3}{2}}v-1\\
\Leftrightarrow v= & \frac{\left(u_{0}^{3}-1\right)}{2}\left(\frac{X}{u_{0}}\right)^{\frac{3}{2}}\pm\sqrt{1-\frac{\left(u_{0}^{3}-1\right)^{2}}{4}\left(\frac{X}{u_{0}}\right)^{3}}.
\end{align*}
so the time, to keep it increasing with $X$, follows
\begin{align*}
t= & \frac{2}{3A}\ln\left\{ \frac{\left(e^{3At_{0}}-1\right)}{2}\left(\frac{X}{e^{At_{0}}}\right)^{\frac{3}{2}}+\sqrt{1-\frac{\left(e^{3At_{0}}-1\right)^{2}}{4}\left(\frac{X}{e^{At_{0}}}\right)^{3}}\right\} \\
= & \frac{2}{3\sqrt{\Omega_{\Lambda}}H_{0}}\ln\left\{ \vphantom{\sqrt{\frac{\left(e^{3At_{0}}-1\right)^{2}}{4}}}\frac{\left(e^{3\sqrt{\Omega_{\Lambda}}H_{0}t_{0}}-1\right)}{2}\left(\frac{a/a_{0}}{e^{\sqrt{\Omega_{\Lambda}}H_{0}t_{0}}}\right)^{\frac{3}{2}}\right.\\
 & \left.+\sqrt{1-\frac{\left(e^{3\sqrt{\Omega_{\Lambda}}H_{0}t_{0}}-1\right)^{2}}{4}\left(\frac{a/a_{0}}{e^{\sqrt{\Omega_{\Lambda}}H_{0}t_{0}}}\right)^{3}}\right\} ,
\end{align*}
with $t_{0}$ solution of
\begin{align*}
t_{0}= & \frac{2}{3\sqrt{\Omega_{\Lambda}}H_{0}}\ln\left\{ \sinh\left(\frac{3}{2}\sqrt{\Omega_{\Lambda}}H_{0}t_{0}\right)+\sqrt{1-\sinh^{2}\left(\frac{3}{2}\sqrt{\Omega_{\Lambda}}H_{0}t_{0}\right)}\right\} .
\end{align*}
The redshift time dependence, based on calculations of Appendix~\ref{sec:AnisoParam} is given by
\begin{align*}
\left(1+z\right)^{2}\left(1+\epsilon-\epsilon_{0}\right)= & \left(\frac{a}{a_{0}}\right)^{-2}\\
\simeq & \left(1+z\right)^{2}\left(1+10^{-\frac{9}{2}}\left[\left(\frac{a}{a_{0}}\right)^{-\frac{3}{2}}-1\right]\epsilon_{r}\right)
\end{align*}

\begin{align*}
\Leftrightarrow1= & \left(1+z\right)^{2}\left(\left(\frac{a}{a_{0}}\right)^{2}\left(1-10^{-\frac{9}{2}}\epsilon_{r}\right)+10^{-\frac{9}{2}}\epsilon_{r}\left(\frac{a}{a_{0}}\right)^{\frac{1}{2}}\right)
\end{align*}
\begin{align*}
\Sigma_{1}= & e^{-2\sqrt{\Omega_{\Lambda}}H_{0}\left(t-t_{0}\right)}\left[\frac{e^{3\sqrt{\Omega_{\Lambda}}H_{0}t}-1}{e^{3\sqrt{\Omega_{\Lambda}}H_{0}t_{0}}-1}\right]^{\frac{4}{3}}\left(1-10^{-\frac{9}{2}}\epsilon_{r}\right),\\
\Sigma_{2}= & 10^{-\frac{9}{2}}\epsilon_{r}e^{-\frac{\sqrt{\Omega_{\Lambda}}H_{0}}{2}\left(t-t_{0}\right)}\left[\frac{e^{3\sqrt{\Omega_{\Lambda}}H_{0}t}-1}{e^{3\sqrt{\Omega_{\Lambda}}H_{0}t_{0}}-1}\right]^{\frac{1}{3}}\\
1+z= & \frac{1}{\sqrt{\Sigma_{1}+\Sigma_{2}}}.
\end{align*}

\section{Luminosidy distance integral}\label{sec:LumDistInt}
The luminosity distance is evaluated by
\begin{align}
a_{0}R\simeq & \left(1-\frac{\epsilon}{2}\right)\int_{t}^{t_{0}}\frac{a_{0}}{a}dt\nonumber \\
= & \left(1-\frac{\epsilon}{2}\right)\frac{e^{-H_{0}t_{0}\sqrt{\Omega_{\Lambda}}}}{H_{0}\sqrt{\Omega_{\Lambda}}}\left(e^{3H_{0}t_{0}\sqrt{\Omega_{\Lambda}}}-1\right)^{\frac{2}{3}}\nonumber \\
 & \times\int_{t>0}^{t_{0}}\frac{e^{H_{0}t\sqrt{\Omega_{\Lambda}}}}{\left(e^{3H_{0}t\sqrt{\Omega_{\Lambda}}}-1\right)^{\frac{2}{3}}}H_{0}\sqrt{\Omega_{\Lambda}}dt.\label{eq:a0R}
\end{align}
Assuming $H_{0}\sqrt{\Omega_{\Lambda}}\equiv A,e^{At}=u,Ae^{At}dt=du$,
we can transform the integral into
\begin{align*}
a_{0}R\simeq & \left(1-\frac{\epsilon}{2}\right)\frac{e^{-At_{0}}}{A}\left(e^{3At_{0}}-1\right)^{\frac{2}{3}}\int_{e^{At}>1}^{e^{At_{0}}}\frac{du}{\left(u^{3}-1\right)^{\frac{2}{3}}}.
\end{align*}
With the variable change $X=u^{3}-1,u=\left(X+1\right)^{\frac{1}{3}},dX=3u^{2}du,du=\frac{dX}{3\left(X+1\right)^{\frac{2}{3}}}$,
we obtain
\begin{align*}
a_{0}R\simeq & \left(1-\frac{\epsilon}{2}\right)\frac{e^{-At_{0}}}{3A}\left(e^{3At_{0}}-1\right)^{\frac{2}{3}}\int_{e^{3At}-1>0}^{e^{3At_{0}}-1}\frac{dX}{\left(X+1\right)^{\frac{2}{3}}X^{\frac{2}{3}}}.
\end{align*}
This can then be integrated, using Mathematica, into
\begin{align*}
a_{0}R\simeq & \left(1-\frac{\epsilon}{2}\right)\frac{e^{-At_{0}}}{3A}\left(e^{3At_{0}}-1\right)^{\frac{2}{3}}\left[3X^{\frac{1}{3}}~{}_{2}F_{1}\left(\frac{1}{3};\frac{2}{3};\frac{4}{3};-X\right)\right]_{e^{3At}-1>0}^{e^{3At_{0}}-1}\\
= & \left(1-\frac{\epsilon}{2}\right)\frac{e^{-At_{0}}}{A}\left(e^{3At_{0}}-1\right)^{\frac{2}{3}}\times\\
&\left[\left(e^{3At}-1\right)^{\frac{1}{3}}~{}_{2}F_{1}\left(\frac{1}{3};\frac{2}{3};\frac{4}{3};1-e^{3At}\right)\right]_{t>0}^{t_{0}},
\end{align*}
so the luminosity distance finally reads
\begin{align*}
a_{0}R\simeq & \left(1-\frac{\epsilon}{2}\right)\frac{e^{-At_{0}}}{A}\left\{ \left(e^{3At_{0}}-1\right)~{}_{2}F_{1}\left(\frac{1}{3};\frac{2}{3};\frac{4}{3};1-e^{3At_{0}}\right)\right.\\
 & \phantom{f}\qquad\left.-\left(e^{3At_{0}}-1\right)^{\frac{2}{3}}\left(e^{3At}-1\right)^{\frac{1}{3}}~{}_{2}F_{1}\left(\frac{1}{3};\frac{2}{3};\frac{4}{3};1-e^{3At}\right)\right\} \\
= & \left(1-\frac{\epsilon}{2}\right)\frac{e^{-H_{0}\sqrt{1-\Omega_{m}}t_{0}}}{H_{0}\sqrt{1-\Omega_{m}}}\left\{ \left(e^{3H_{0}\sqrt{1-\Omega_{m}}t_{0}}-1\right)\right.\\
 & \phantom{f}\qquad\qquad~{}_{2}F_{1}\left(\frac{1}{3};\frac{2}{3};\frac{4}{3};1-e^{3H_{0}\sqrt{1-\Omega_{m}}t_{0}}\right)\\
 & -\left(e^{3H_{0}\sqrt{1-\Omega_{m}}t_{0}}-1\right)^{\frac{2}{3}}\left(e^{3H_{0}\sqrt{1-\Omega_{m}}t}-1\right)^{\frac{1}{3}}\\
 & \phantom{f}\qquad\qquad\hfill\left.~{}_{2}F_{1}\left(\frac{1}{3};\frac{2}{3};\frac{4}{3};1-e^{3H_{0}\sqrt{1-\Omega_{m}}t}\right)\right\} .
\end{align*}
The indefinite integral from Eq.~\eqref{eq:a0R} 
indeed is positive and increasing, as expected.

\section{Anisotropy parameter}\label{sec:AnisoParam}

Using \cite[Eqs.~(A4) and (15)]{LeDelliou:2020kbm}, we can relate the anisotropy parameter with its value at recombination as
\begin{align*}
\frac{\epsilon}{\epsilon_{r}}= & \frac{\sqrt{\Omega_{m}\left(\frac{a_{i}}{a_{0}}\right)^{-3}+1-\Omega_{m}}-\sqrt{\Omega_{m}\left(\frac{a}{a_{0}}\right)^{-3}+1-\Omega_{m}}}{\sqrt{\Omega_{m}\left(\frac{a_{i}}{a_{0}}\right)^{-3}+1-\Omega_{m}}-\sqrt{\Omega_{m}\left(\frac{a_{r}}{a_{0}}\right)^{-3}+1-\Omega_{m}}}\\
= & \frac{1-\sqrt{\frac{\left(\frac{a}{a_{0}}\right)^{-3}+\frac{1}{\Omega_{m}}-1}{\left(\frac{a_{i}}{a_{0}}\right)^{-3}+\frac{1}{\Omega_{m}}-1}}}{1-\sqrt{\frac{\left(\frac{a_{r}}{a_{0}}\right)^{-3}+\frac{1}{\Omega_{m}}-1}{\left(\frac{a_{i}}{a_{0}}\right)^{-3}+\frac{1}{\Omega_{m}}-1}}}.
\end{align*}
Since we consider $\Omega_{m}\sim1$, $\frac{a_{r}}{a_{0}}\sim10^{-3}\ll1$
and $\frac{a_{i}}{a_{r}}\sim10^{n}\gg1$ from \cite[Eq.~(15)]{LeDelliou:2020kbm},
then
\begin{align*}
\frac{\epsilon}{\epsilon_{r}}\simeq & \frac{1-\sqrt{\frac{\left(\frac{a}{a_{0}}\right)^{-3}+\frac{1}{\Omega_{m}}-1}{\left(\frac{a_{i}}{a_{0}}\right)^{-3}}}}{-\left(\frac{a_{r}}{a_{i}}\right)^{-\frac{3}{2}}}\\
\simeq & \left(\frac{a_{r}}{a_{0}}\right)^{\frac{3}{2}}\sqrt{\left(\frac{a}{a_{0}}\right)^{-3}+\frac{1}{\Omega_{m}}-1}-\left(\frac{a_{r}}{a_{i}}\right)^{\frac{3}{2}}\\
\simeq & 10^{-\frac{9}{2}}\sqrt{\left(\frac{a}{a_{0}}\right)^{-3}+\frac{1}{\Omega_{m}}-1}-10^{-\frac{3}{2}n}\\
\simeq & 10^{-\frac{9}{2}}\left(\sqrt{\left(\frac{a}{a_{0}}\right)^{-3}+\frac{1}{\Omega_{m}}-1}-10^{-\frac{3}{2}\left(n-3\right)}\right)>0
\end{align*}
for
\begin{align*}
\left(\frac{a}{a_{0}}\right)^{-3}> & 10^{-3\left(n-3\right)}+1-\frac{1}{\Omega_{m}}\simeq10^{-3\left(n-3\right)}\\
\Leftrightarrow\frac{a}{a_{0}}< & 10^{\left(n-3\right)},n\sim2\Rightarrow a<10^{-1}a_{0}
\end{align*}
so setting $\frac{a}{a_{0}}=10^{-\left(1+\eta\right)}\Leftrightarrow\eta=-1-\log\frac{a}{a_{0}}$
\begin{align*}
\frac{\epsilon}{\epsilon_{r}}\simeq & 10^{-\frac{9}{2}}\left(\left(\frac{a}{a_{0}}\right)^{-\frac{3}{2}}-10^{-\frac{3}{2}\left(n-3\right)}\right)\\
\simeq & 10^{-\frac{9}{2}}10^{\frac{3}{2}\left(1+\eta\right)}\left(1-10^{-\frac{3}{2}\left(n-2+\eta\right)}\right)\\
\simeq & 10^{\frac{3}{2}\left(\eta-2\right)}=10^{-\frac{3}{2}\left(\log\frac{a}{a_{0}}+3\right)}=10^{-\frac{9}{2}}\left(\frac{a}{a_{0}}\right)^{-\frac{3}{2}}
\end{align*}


\bsp	
\label{lastpage}
\end{document}